\documentclass[aip,reprint,eqseqnum,bibnotes,showpacs,superscriptaddress]{revtex4-1}
\usepackage{amsmath,amsthm,amsfonts,amssymb,bm}
\usepackage{graphicx}
\usepackage{color}
\usepackage{natbib}
\usepackage{mciteplus}

\newcommand{\base}[1]{{#1}_0}
\newcommand{\gdot}{\dot{\gamma}}
\newcommand{\gdotbar}{\overline{\dot{\gamma}}}

\newcommand{\beqn}{\begin{equation}}
\newcommand{\eeqn}{\end{equation}}
\newcommand{\beqna}{\begin{eqnarray}}
\newcommand{\eeqna}{\end{eqnarray}}

\newcommand{\taur}{\tau_R}
\newcommand{\taud}{\tau_d}
\newcommand{\tauk}{\tau_k}

\newcommand{\etal}{\emph{et al. }}
\newcommand{\invtaur}{\taur^{-1}}

\newcommand{\invtaud}{\taud^{-1}}

\newcommand{\deltag}{\delta \gdot_n}
\newcommand{\gdotsrp}{\gdot_{0,s}}
\newcommand{\gdotnrp}{\gdot_{0,n}}

\newcommand{\vecv}[1]{\bm{{#1}}}
\newcommand{\tens}[1]{\bm{{#1}}}

\newcommand{\visc}{\tens{W}} 

\newcommand{\wxy}{W_{xy}}
\newcommand{\wyy}{W_{yy}}
\newcommand{\wxx}{W_{xx}}

\newcommand{\vs}{\emph{vs}.\ }

\usepackage{graphicx,amsmath,color,amsfonts}

\begin{document}
 
\title{Shear banding in time-dependent flows of polymers and wormlike micelles}

\author{R. L. Moorcroft} 
\author{S. M. Fielding} 

\affiliation{Department of Physics, Durham University, Science
  Laboratories, South Road, Durham, DH1 3LE, UK}

\date{\today}
\begin{abstract} { We study theoretically the formation of shear bands
    in time-dependent flows of polymeric and wormlike micellar
    surfactant fluids, focussing on the protocols of step shear
    stress, step shear strain (or in practice a rapid strain ramp),
    and shear startup, which are commonly studied experimentally.  For
    each protocol we perform a linear stability analysis to provide a
    fluid-universal criterion for the onset of shear banding,
    following our recent Letter~\cite{Moorcroftetal2012}. In each case
    this criterion depends only on the shape of the experimentally
    measured rheological response function for that protocol,
    independent of the constitutive properties of the material in
    question. (Therefore our criteria in fact concern all complex
    fluids and not just the polymeric ones of interest here. A
    separate manuscript~\cite{SuzanneInProgress} will explore them in
    a broad class of disordered soft glassy materials including foams,
    dense emulsions, dense colloids, and microgel bead suspensions.)
    An important prediction is that pronounced banding can arise
    transiently in each of these protocols, even in fluids for which
    the underlying constitutive curve of stress as a function of
    strain-rate is monotonic and a steadily flowing state is
    accordingly unbanded.  For each protocol we provide numerical
    results in the rolie-poly and Giesekus models that support our
    predictions. We comment on the ability of the rolie-poly model to
    capture the observed experimental phenomenology, and on the
    failure of the Giesekus model.}
\end{abstract}
\maketitle

\section{Introduction}

\newif\iffigures
\figurestrue 

Many complex fluids show shear banding, in which an initially
homogeneous shear flow undergoes an instability leading to the
formation of macroscopic bands of differing viscosity, which coexist
at a common shear stress \cite{Ovarlezetal2009a, Manneville2008a,
  Olmsted2008a, Fielding2007a}.  Examples include entangled polymer
solutions and melts \cite{Wangetal2008a,Wangetal2006b,Wangetal2009a,
  Wangetal2009d}, triblock copolymer solutions
\cite{Berretetal2001a,Mannevilleetal2007a}, wormlike micellar
surfactant solutions
\cite{Lerougeetal2010a,Wangetal2008c,Helgesonetal2009a,Helgesonetal2009b,Huetal2008a,Milleretal2007a,
  Salmonetal2003b,Mairetal1997a,Mairetal1996a,Makhloufietal1995a},
lyotropic lamellar surfactant phases \cite{Salmonetal2003a},
concentrated suspensions and emulsions
\cite{Coussotetal2002c,Paredesetal2011a}, carbopol microgels
\cite{Divouxetal2010a}, star polymers \cite{Rogersetal2008a} and
foams \cite{Rodtsetal2005a}.

To date, most studies have focused on the long-time rheological
response of these fluids, once a steady flowing state has been
established.  The criterion for shear banding in this steady state
limit is well known~\cite{Yerushalmietal1970a}: that there exists a
region of negative slope in the constitutive curve of shear stress as
a function of shear rate for an underlying base state of stationary
homogeneous flow. The steady state flow curve relation between stress
and strain rate then exhibits a characteristic plateau in the shear
banding regime, signifying a coexistence of bands of differing shear
rates $\gdot_{\rm l},\gdot_{\rm h}$ at a common value $\Sigma_{\rm p}$
of the shear stress. See Fig.~\ref{fig: flow_curve}.

However most practical flows involve a strong time-dependence, whether
perpetually or during the initial startup of deformation before a
steadily flowing state has been established. Accordingly, increasing
experimental attention is now being devoted to time-dependent flow
protocols. Shear banding has recently been reported following the
imposition of a step stress~\cite{Gibaudetal2010a,Divouxetal2011b,
  Wangetal2009a,Huetal2007a, Wangetal2003a, Huetal2008a,
  Wangetal2008c, Huetal2005a, Huetal2010a}, following a step shear
strain~\cite{Wangetal2010a, Wangetal2009c, Wangetal2006a,
  Fangetal2011a, Wangetal2007a, Archeretal1995a, Wangetal2008c}, and
during shear startup~\cite{Divouxetal2010a, Divouxetal2011a,
  Wangetal2009a, Huetal2007a, Wangetal2008a}.

In each case the onset of banding appears closely linked to the
presence of a distinctive signature in the shape of the material's
time-dependent rheological response function. Importantly, although
this signature is specific to the particular flow protocol in
question, it appears largely universal for all complex fluids in a
given protocol.  For example the onset of shear banding in the shear
startup protocol appears closely related to the presence of an
overshoot in the stress startup signal, as we shall elaborate below.

Motivated by these observations, in a recent
Letter~\cite{Moorcroftetal2012} we derived fluid-universal criteria
for the onset of linear instability to the formation shear bands, one
for each protocol in turn: step stress, shear startup, and step
strain. Each criterion depends only on the shape of the experimentally
measured rheological response function for that protocol, {\em
  independent} of the mesostructure and constitutive dynamics of the
particular material in question. These predictions for banding in
time-dependent flows thus have the same highly general,
fluid-universal status as the widely known criterion for banding in
steady state (of a negatively sloping constitutive curve).

Whether or not the time-dependent shear bands predicted here persist
to steady state of course depends on the shape of that underlying
constitutive curve for stationary homogeneous flow.  However an
important contribution of this work is to highlight that pronounced
banding often arises during a fluid's transient evolution to a steady
flowing state, given the time-dependent flow signatures that we shall
discuss, even in fluids for which the underlying constitutive curve is
monotonic and the eventual steady state unbanded.

The present manuscript provides an in-depth discussion of the criteria
outlined in Ref.~\cite{Moorcroftetal2012}, and a thorough numerical
exploration of them within two of the most popular models for the
rheology of entangled polymeric fluids: the rolie-poly (RP) model and
the Giesekus model.  Accordingly, it addresses conventional polymeric
fluids such as concentrated polymer solutions or melts of high
molecular weight; as well as entangled wormlike micelles whose long,
chain-like substructures undergo the same stress relaxation mechanisms
as polymers, with the additional mechanisms of chain breakage and
reformation~\cite{Cates1990a}. For convenience we refer to all these
materials simply as `polymeric fluids' in what follows. The reader is
referred to a separate manuscript~\cite{SuzanneInProgress} for a
discussion of the same phenomena in the context of a broad class of
disordered soft glassy materials such as foams, dense emulsions, onion
surfactants and microgel bead suspensions.

The paper is structured as follows.  In section~\ref{sec:observations}
we survey the experimental and simulation evidence for shear banding
in time-dependent flow protocols.  In Sec.~\ref{section: models} we
outline a general theoretical framework for the rheology of complex
fluids, and give details of the rolie-poly and Giesekus constitutive
models of polymeric flows.  In Sec.~\ref{section: stability_analysis}
we detail a linear stability analysis for the onset of shear banding
in time-dependent flows, performed within this general framework.  In
Secs.~\ref{section: stepstress},~\ref{section: rampstrain}
and~\ref{section: shearstartup} we present our analytical criteria for
the onset of shear banding in step stress, strain ramp, and shear
startup protocols respectively, and give supporting numerical evidence
in each case. We also discuss the way our predictions relate to
experimental data.  Conclusions and perspectives for further study are
given in Sec.~\ref{section: conclusions}.

\iffigures
\begin{figure}[tbp]
  \centering
  \includegraphics[width=5cm,angle=270]{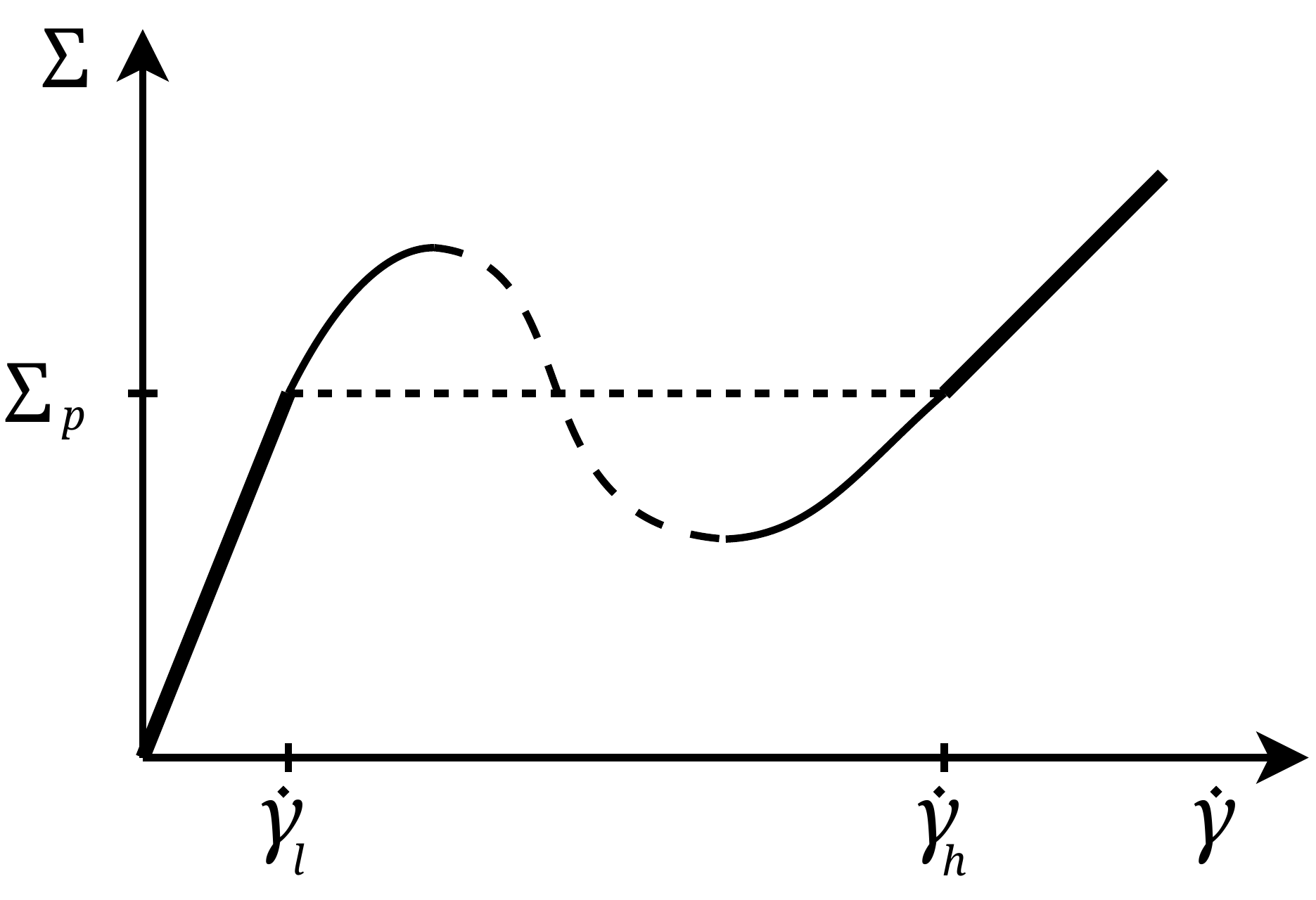}
  \caption{Thin line: constitutive curve for an underlying base state of homogeneous shear flow. Homogeneous flow is linearly unstable in the dashed region. Thick lines joined by dotted plateau: corresponding steady state flow curve. For imposed shear rates in the plateau region $\gdot_{\rm l} < \gdotbar < \gdot_{\rm h}$ the steady state is shear banded: see Fig. \ref{fig: banding}.}
  \label{fig: flow_curve}
\end{figure}
\fi

\iffigures
\begin{figure}[tbp]
  \centering
  \includegraphics[width=3cm,angle=270,trim=5cm 0cm 0cm 0cm, clip=true]{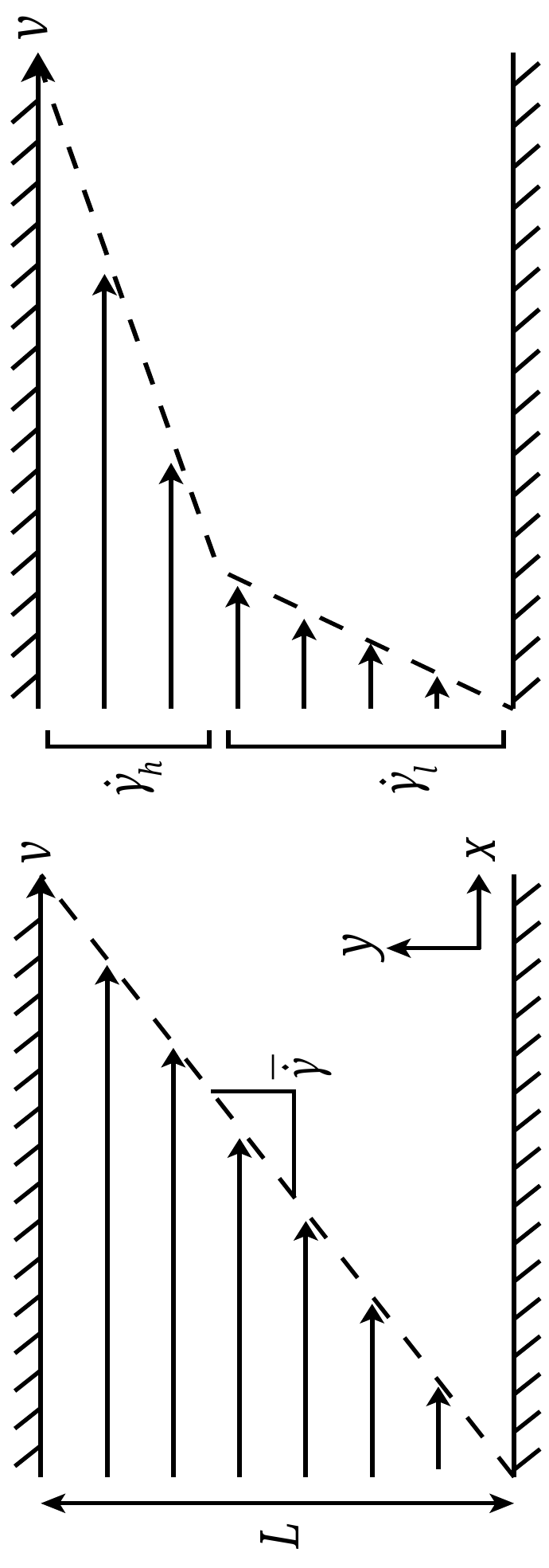}
  \caption{{\bf Left:} homogeneous flow profile. {\bf Right:} shear banded flow profile.}
  \label{fig: banding}
\end{figure}
\fi

Throughout the manuscript we use the term `constitutive curve' to
describe the relation between shear stress and strain rate for an
underlying base state of homogeneous shear flow. We use the term `flow
curve' to describe the relation between shear stress and applied
strain rate $\gdotbar$ (averaged across the sample) for a steady
flowing state.  In any regime of homogeneous steady state flow, these
two curves coincide.

\section{Experimental and numerical motivation}
\label{sec:observations}

\subsection{Step stress}

Steady state shear banding is associated with a region of negative
slope in the underlying constitutive curve of shear stress $\Sigma$ as
a function of shear rate $\gdot$ for an underlying base state of
stationary (though not necessarily stable) homogeneous flow. The
composite flow curve $\Sigma(\gdot)$ for the steady state banded flow
then typically displays a flat (or slightly upwardly sloping) plateau
spanning a window of shear rates $\gdot_{\rm l} <\gdot < \gdot_{\rm
  h}$ at (or spanning a small window about) a selected value of the
stress $\Sigma_{\rm p}$. See Fig.~\ref{fig: flow_curve}. Accordingly,
steady state banding is relatively easily accessed in an applied shear
rate protocol for imposed rates $\gdot_{\rm l} <\gdot < \gdot_{\rm
  h}$, but much more difficult to access under conditions of a
constant imposed stress, which must be tuned to lie in the small
window of stress consistent with this near-plateau in the flow curve.

Nonetheless, following the imposition to a previously unloaded sample
of a step stress in the vicinity of this plateau $\Sigma \sim
\Sigma_p$, entangled polymeric materials commonly exhibit
time-dependent shear banding
\cite{Wangetal2009a,Huetal2007a,Wangetal2003a, Huetal2008a,
  Wangetal2008c, Huetal2005a, Huetal2010a, Wangetal2009d}.  Its onset
appears closely associated with a sudden and dramatic increase in the
shear rate response of the system by several orders of magnitude over
a short time interval, as it rises from a small initial value to
attain its final value on the steady state flow curve
\cite{Wangetal2008b, Wangetal2003a, Huetal2008a, Wangetal2008c,
  Huetal2005a, Wangetal2009d}.  In some cases a return to homogeneous
shear is seen as steady state is neared \cite{Wangetal2009a}, though
this is often complicated by the occurrence of edge fracture, which
can severely limit the determination of true steady state
\cite{Wangetal2008b,Innetal2005a}.  Whether polymeric fluids exhibit
steady state banding in the step stress protocol therefore remains an
open question. 

\subsection{Strain ramp}

In a strain ramp protocol, shear is applied at a constant rate
$\gdot_0$ for a time $t^*$ until some desired strain amplitude
$\gamma_0 = \gdot_0 t^*$ is attained, after which the shearing is
stopped.  The limit $t^*\to 0$ and $\gdot_0 \to \infty$ at fixed
$\gamma_0$ gives a theoretically idealized step strain. Indeed in
practice a rapid ramp is often termed a step strain.

The shear stress relaxation function $\Sigma(t')$ as a function of the
time $t' = t-t^*$ elapsed since the end of the ramp is usually
reported after scaling by the strain amplitude to give $G(t',\gamma_0)
= \Sigma(t',\gamma_0)/\gamma_0$, with $G(t') = \lim_{\gamma_0 \to 0}
G(t',\gamma_0)$ in the small strain limit of linear response.

This protocol has been extensively studied experimentally in entangled
polymeric fluids~\cite{Venerus2005a, Einagaetal1971a, Osakietal1980a,
  Osakietal1982a, Larson1988, DoiEdwards, Osaki1993a, Einagaetal1970a,
  Garridoetal2009a, Sanchez-Reyesetal2002a}.  The stress relaxation
function typically shows a double exponential form, with time-strain
separability characterised by the so-called ``damping function'' $h$
for times greater than some $t' = \tauk$:
\beqn
h(\gamma_0) = \frac{G(t',\gamma_0)}{G(t')} \quad \text{ for } t' \gg \tauk.
\label{eqn: damping_fn}
\eeqn

Experimental data for this damping function in entangled polymers has
been compared extensively with the form predicted theoretically by Doi
and Edwards (DE)~\cite{DE2, Osakietal1982a, Garridoetal2009a,
  Sanchez-Reyesetal2002a, Osakietal1980a}.  Review articles
\cite{Venerus2005a, Osaki1993a} suggest that a significant proportion
of the available experimental data \cite{Osakietal1982a, Archer1999a,
  Islametal2001a, Vrentasetal1982} agrees well with the DE damping
function, particularly for moderately entangled melts or solutions.
This has been termed type A behaviour.  A small number of studies,
mostly in very weakly entangled fluids, show a weaker initial stress
relaxation in the regime before the time-strain separable domain is
reached, leading to a damping function that lies above the Doi-Edwards
prediction (type B behaviour).  Finally, studies~\cite{Osakietal1980a}
of fluids with high entanglement numbers $Z \gtrsim 50$ often report a
faster stress relaxation before the time-strain separable domain is
reached, so that the experimental damping function lies below that of
Doi and Edwards~\cite{Julianietal2001a, Archer1999a, Islametal2001a,
  Vrentasetal1982} (type C behaviour).

Velocimetric studies in this protocol commonly reveal `macroscopic
motions' following a step strain of sufficiently large amplitude
$\gamma_0 \gtrsim 1.5$ in both entangled polymer melts and solutions
\cite{Wangetal2010a, Wangetal2009c, Wangetal2006a, Wangetal2007a,
  Fangetal2011a, Archeretal1995a} and wormlike micelles
\cite{Wangetal2008c}. These macroscopic motions refer to non-zero
velocities $v(y,t>t^*) \neq 0$ and associated heterogeneous shear
zones -- {\it i.e.}, shear bands -- measured locally within the flow
cell for some time even once the rheometer plates have stopped moving
after the end of the ramp. Wang and co-workers showed that the same
fluid can exhibit all three types of behaviour A to C above, depending
on the extent of any wall slip associated with these `macroscopic
motions'~\cite{Wangetal2007a}: bulk shear in the sample's interior is
usually associated with type A or B behaviour, and wall-slip with type
C behaviour.

Theoretically, an instability leading to strain localisation following
the imposition of a step strain was proposed in the context of the DE
model by Marrucci and Grizzuti \cite{MarrucciGrizzuti1983a}.  The DE
theory predicts a maximum in the shear stress $\Sigma(t',\gamma_0)$
when it is plotted as a function of applied strain amplitude
$\gamma_0$ for a given time instant $t' \sim \tauk$ after the step.
This maximum occurs at a strain value of $\gamma_0 \approx 2$ and
results in a negative slope for strain amplitudes beyond this maximum:
$\partial_{\gamma_0} \Sigma(t'\sim\tauk,\gamma_0) < 0$. Marrucci and
Grizzuti used a free energy calculation to show that step strains with
amplitudes in this regime of negative slope are unstable to the onset
of heterogeneity.

Numerical studies of the rolie-poly model by Olmsted and co-workers
have likewise reported shear rate heterogeneity during stress
relaxation after a fast strain ramp of sufficiently large strain
amplitude~\cite{Adamsetal2009a, Adamsetal2009b} consistent with the
predictions of Ref.~\cite{MarrucciGrizzuti1983a}. The form of this
heterogeneity is sensitive to the initial noise conditions, and its
onset can show a delay after the end of the ramp
\cite{Agimelenetal2012a}. These results are in qualitative accordance
with experiments showing that the onset of macroscopic motions can be
delayed after the end of the ramp \cite{Wangetal2009b, Wangetal2009c,
  Archeretal1995a}, with the delay time related to the time taken for
polymer chain stretch to relax~\cite{Wangetal2009b, Archeretal1995a}.
Olmsted and co-workers also showed that in extreme cases a very large
shear rate can develop across a stationary `fracture' plane, so that
the local velocity is very difficult to resolve. These results are
qualitatively similar to experiments showing a `failure' plane over
which the shear rate is extremely high \cite{Fangetal2011a,
  Wangetal2009b}.  Shear rate heterogeneity after a fast ramp has also
been reported in a two species elastic network
model~\cite{Zhouetal2008a}.

\subsection{Shear startup}

In shear startup, a constant shear rate is applied to a previously
undeformed sample for all times $t>0$.  Time-dependent shear banding
has been widely reported in this protocol in entangled polymer
solutions and melts \cite{Wangetal2008a, Wangetal2008b, Wangetal2009a,
  Wangetal2009b, Wangetal2009c, Wangetal2003a, Huetal2007a,
  Wangetal2009d}, and in wormlike micelles \cite{Huetal2008a,
  Wangetal2008c}.  Its onset appears closely associated with the
presence of an overshoot in the shear stress as it evolves from its
initial value of zero to its final value on the ultimate steady state
flow curve.

In some materials this time-dependent banding is effectively a
precursor to true steady state banding. In such cases it can be viewed
as the kinetic process by which these bands develop out of an
initially homogeneous startup flow.  Nonetheless, the magnitude of the
time-dependent banding during startup is often strikingly greater than
that which remains in the final steady state \cite{Wangetal2009a,
  Wangetal2008b, Wangetal2008a, Wangetal2006b, Wangetal2009d}. Indeed,
it is often sufficiently dramatic as to be accompanied by elastic-like
recoil in which velocities measured locally within the flow cell can
even temporarily become negative~\cite{Wangetal2009a, Wangetal2008a,
  Wangetal2009d} such that the fluid is locally and temporarily moving
in the direction opposite to that of the rheometer plate that is
driving the shear.  Furthermore, pronounced but transient banding
associated with a stress overshoot is also commonly seen even in less
well entangled polymer solutions, for which the final steady flow
state is homogeneous \cite{Wangetal2009a, Huetal2007a, Wangetal2008a}.
Taken together, this evidence suggests that qualitatively different
instability mechanisms might underlie shear banding in startup
compared with that in steady state. We return to explore this concept
in our discussion of `elastic' versus `viscous' banding instabilities
in Sec.~\ref{section: shearstartup} below.

Numerical studies have likewise reported time-dependent shear banding
associated with stress overshoot in startup. Refs.
\cite{Adamsetal2009a, Adamsetal2009b, Adamsetal2011} explored the
rolie-poly model of polymeric fluids, which can have either a
monotonic or nonmonotonic constitutive curve, depending the value of
the convective constraint release parameter $\beta$ and the
entanglement number $Z$.  This work demonstrated that banding and
negative-velocity recoil can arise shortly after a stress overshoot,
regardless of whether the underlying constitutive curve is monotonic
or nonmonotonic.  For a nonmonotonic constitutive curve this banding
persists to steady state, though with a much weaker magnitude than
during startup.  For a monotonic constitutive curve, homogeneous flow
is recovered in steady state. Ref.~\cite{Adamsetal2011} also discussed
carefully the effects of rheometer cell curvature on these phenomena.
Banding in startup has also been reported in simulations of a
two-species elastic network model \cite{Zhouetal2008a}; and in
molecular dynamics simulations of polymer
melts~\cite{Likhtmanetal2012}.


\section{Models}
\label{section: models}

\subsection{Force balance}

The stress response of a complex fluid to an applied deformation is
dominated by the behaviour of its internal mesoscopic substructures
\cite{Larson1999}. For example, a polymeric fluid comprises many
chain-like molecules, the entanglements between which result in
topological constraints on their molecular motion. We therefore
decompose the total stress $\tens{\Sigma}$ into a viscoelastic
contribution from these mesoscopic substructures, as well as a
familiar Newtonian contribution of viscosity $\eta$, and an isotropic
pressure:
\beqn
\tens{\Sigma} = G(\visc -\tens{I})+ 2 \eta \tens{D} - p\tens{I}.
\label{eqn: total_stress_tensor}
\eeqn

The viscoelastic contribution $\tens{\sigma} = G\, (\visc - \tens{I})$
is expressed in terms of a constant elastic modulus $G$ and a tensor
$\visc(\tens{r},t)$, which describes the local conformation of the
mesoscopic substructures.  The dynamics of this conformation tensor in
flow is prescribed by a viscoelastic constitutive equation. In the
next section below we introduce the constitutive equations that we
shall use throughout this work.  The Newtonian contribution $2 \eta
\tens{D}$ may arise from the presence of a true a solvent, or may
represent viscous stresses arising from any fast degrees of freedom of
the polymer chains that are not ascribed to the viscoelastic
contribution. Here $\tens{D} = \frac{1}{2}(\tens{K} + \tens{K}^T)$
where $K_{\alpha\beta} =
\partial_{\beta}v_{\alpha}$ and $\tens{v}(\tens{r},t)$ is the fluid
velocity field.  The isotropic pressure field $p(\tens{r},t)$ is
determined by the condition of incompressible flow:
\beqn
\label{eqn: incomp}
\vecv{\nabla}\cdot\vecv{v} = 0.
\eeqn

Throughout we consider the limit of zero Reynolds number, in which the
force balance condition states that the stress tensor $\tens{\Sigma}$
must remain divergence free:
\beqn
\vecv{\nabla}\cdot\,\tens{\Sigma} = 0.
\label{eqn: force_balance}
\eeqn

\subsection{Viscoelastic constitutive equation}

The rheological response of an entangled polymeric fluid can be
modelled from a microscopic starting point by considering a test
polymer chain that has its dynamics laterally constrained by
topological entanglements with other chains.  These entanglements are
then represented in mean field spirit by an effective `tube'
\cite{DoiEdwards}.  The GLAMM model \cite{GLAMM} provides a fully
microscopic stochastic equation of motion for such a test chain and
its tube. However it is computationally intensive to work with in
practice.  An approximation was therefore derived in
Ref.~\cite{Grahametal2003a} by projecting the full GLAMM model onto a
single mode description, known as the rolie-poly (RP) model. This
gives the viscoelastic constitutive equation for the dynamics of the
conformation tensor as:\\
\begin{widetext}
\beqna
\partial_t{\visc}+\tens{v}.\nabla\tens{\visc} &=& \tens{K} \cdot \visc + \visc \cdot \tens{K}^T - \frac{1}{\taud}\left(\visc - \tens{I}\right) 
- \frac{2(1-A)}{\taur}\left[\, \visc + \beta A^{-2\delta}\left(\visc - \tens{I}\right) \right] + D\nabla^2\visc.
\label{eqn: rolie-poly_tensor}
\eeqna
\end{widetext}
Here $A = \sqrt{3/T\,}$, where $T = \text{tr}\,\tens{\visc}$ denotes
the magnitude of chain stretch. The reptation time $\taud$ is the
timescale on which a test chain escapes its tube of constraints by
undergoing 1D curvilinear diffusion along its own length. The Rouse
time $\taur$ is the much shorter timescale on which chain stretch
relaxes.  The ratio of these two relaxation times is prescribed by the
number of entanglements per chain \cite{DoiEdwards} $Z$, with $
\taud/\taur=3Z$.  The parameter $\beta$ describes the efficacy of
so-called convective constraint release (CCR) events, in which
relaxation of polymer chain stretch also relaxes entanglement points,
thereby also allowing relaxation of tube orientation. It has range $0
\leq \beta \leq 1$.  The parameter $\delta$ also describes CCR.
Following Ref.~\cite{Grahametal2003a} we set $\delta = -\frac{1}{2}$
throughout.  Depending on the values of the model parameters, the
constitutive curve of the RP model can either be monotonic or
non-monotonic.

The value of $\beta$ is difficult to relate directly to experiment and
there is no consensus on its correct value, though a small value was
used by Likhtman and Graham to best fit experimental
data~\cite{Grahametal2003a} in flow protocols that were assumed to be
homogeneous.  A more recent study \cite{Agimelenetal2012a} aimed at
describing `fracture-like' velocity profiles after a step strain
likewise found a good fit to experimental findings only for small
values of $\beta$.

The diffusive term $D\nabla^2\visc$ was absent in the original
formulation of the model.  Without it, however, the interface between
the shear bands is unphysically sharp, with a discontinuity in the
shear rate profile $\gdot(y)$ across it.  Furthermore the total shear
stress of a steady shear banded state is not uniquely selected, but
depends on the shear history to which the material has been
subject~\cite{Olmstedetal2000a}. This contradicts experimental
findings, which find a unique plateau stress $\Sigma_{\rm p}$.  The
diffusive term lifts this degeneracy to give a uniquely selected
stress as well as a characteristic width to the interface between the
bands of $\ell =
O(\sqrt{D\taud})$~\cite{Luetal2000a,Olmstedetal2000a}.

A more phenomenologically motivated constitutive equation for
concentrated polymeric solutions or melts considers an anisotropic
drag on polymer chains that are oriented due to flow.  Representing
these chains simply as dumbells, Giesekus~\cite{Giesekus1982} began
with the upper convected Maxwell model for dilute solutions and
incorporated into it an anisotropy parameter $\alpha$ with $0\leq
\alpha \leq 1$.  The resulting constitutive equation has the form:\\
\begin{widetext}
\beqna
\partial_t{\visc}+\tens{v}.\nabla\tens{\visc} 
&=& \tens{K}\cdot \visc + \visc \cdot \tens{K}^T - \frac{1}{\lambda}\left(\visc - \tens{I}\right) - \frac{\alpha}{\lambda}\left(\visc - \tens{I}\right)^2+D\nabla^2\visc,
\label{eqn: giesekus_constit}
\eeqna
\end{widetext}
where $\lambda$ is the relaxation time. A diffusive term is again
included to properly describe a shear banding flow.

This Giesekus model admits either nonmonotonic or monotonic
constitutive curves, depending on the value of $\alpha$ and the
solvent viscosity $\eta$. It has been successful in modelling the
steady state shear banding properties of entangled wormlike micelles
\cite{Helgesonetal2009a,Helgesonetal2009b}, and a multimode equivalent
has shown good agreement with the experimentally measured steady shear
viscosity
\cite{Byarsetal1997a,Quinzanietal1990,Burdette1989,Azaiezetal1996} and
damping function \cite{Khanetal1987} of polymeric materials.

\subsection{Flow geometry}

Throughout we consider a sample of fluid sandwiched between parallel
plates at $y = \{0,L\}$, sheared by moving the top plate in the
$\vecv{\hat{x}}$ direction. Translational invariance is assumed in the
$\vecv{\hat{x}},\vecv{\hat{z}}$ directions. The fluid velocity is then
of the form $\vecv{v} = v(y,t)\vecv{\hat{x}}$, and the local shear
rate
\beqn
\gdot(y,t) = \partial_{y}v(y,t).
\eeqn
The spatially averaged (or `global') shear rate is
\beqn
\gdotbar(t) = \frac{1}{L}\int_{0}^{L} \gdot(y,t)dy.
\eeqn

\subsection{Componentwise equations}

In the flow geometry just described, the condition of incompressible
flow (Eqn.~\ref{eqn: incomp}) is automatically satisfied.  The force
balance condition of creeping flow (Eqn.~\ref{eqn: force_balance})
demands that the total shear stress is uniform across the cell
$\partial_{y}\Sigma_{xy} =0$.  The viscoelastic and Newtonian solvent
contributions may however each vary in space, provided their sum
remains uniform:
\beqn
\Sigma_{xy}(t) = G\wxy(y,t) + \eta \gdot(y,t).
\label{eqn: shear_stress}
\eeqn

Componentwise the RP model reduces to a system of three dynamical
variables:\\
\begin{widetext}
\beqna
\dot{W}_{xy}  &=& \gdot \wyy - \frac{\wxy}{\taud} - \frac{2(1-A)}{\taur}(1+ \beta A)\wxy + D\partial_y^2 \wxy, \nonumber\\
\dot{W}_{yy}  &=& - \frac{\wyy-1}{\taud} - \frac{2(1-A)}{\taur}\left[\wyy+ \beta A(\wyy-1)\right]+ D\partial_y^2\wyy,\nonumber\\
\dot{T}     &=&  2\dot{\gamma}\wxy  - \frac{T-3}{\taud} - \frac{2(1-A)}{\taur}\left[T + \beta A(T - 3)\right]+ D\partial_y^2 T.\quad \quad
\label{eqn: sRP_components}
\eeqna
In the limit of fast chain stretch relaxation $\taur \to 0$ this reduces to
a simpler system of two dynamical variables:
\beqna
\dot{W}_{xy} &=& \dot{\gamma} \left[\wyy - \frac{2}{3} (1+\beta)\wxy^2\right]\;\;\;\;\;\;\; -\frac{1}{\taud}\wxy,+ D\partial_y^2 \wxy\nonumber\\
\dot{W}_{yy} &=& \frac{2}{3}\dot{\gamma}\left[\beta\wxy-(1+\beta)\wxy\wyy \right] - \frac{1}{\taud}(\wyy-1)+ D\partial_y^2\wyy. \quad \quad
\label{eqn: nRP_components}
\eeqna
\end{widetext}
with a constant molecular trace $T=3$. We shall refer to this
`non-stretching' form below as the nRP model; and the full
`stretching' version of Eqns.~\ref{eqn: sRP_components} as the sRP
model.

The Giesekus model likewise reduces to a system of three dynamical variables:\\
\begin{widetext}
\beqna
\dot{W}_{xy} &=& \gdot \wyy \,\,\,- \frac{\wxy}{\lambda} - \frac{\alpha\wxy}{\lambda}\left[(\wxx - 1) + (\wyy - 1) \right]+ D\partial_y^2 \wxy, \nonumber \\
\dot{W}_{xx} &=& 2\wxy\gdot - \frac{\wxx-1}{\lambda} - \frac{\alpha}{\lambda}\left[ \wxy^2 + (\wxx - 1)^2\right]+ D\partial_y^2\wxx, \nonumber \\
\dot{W}_{yy} &=& \quad \quad \quad \, -\frac{\wyy -1}{\lambda} \, - \, \frac{\alpha}{\lambda}\left[\wxy^2 + (\wyy-1)^2 \right]+ D\partial_y^2\wyy.
\label{eqn: Giesekus_components}
\eeqna
\end{widetext}

We note an important distinction in the structure of these equations.
In particular, in the stretching rolie-poly model Eqns.~\ref{eqn:
  sRP_components} and the Giesekus model Eqns.~\ref{eqn:
  Giesekus_components}, the terms prefactored by $\gdot$ are of simple
linear form, whereas the terms prefactored by the inverse relaxation
timescales are nonlinear. Conversely in the nonstretch rolie-poly
model the terms prefactored by $\gdot$ are nonlinear, whereas the
terms prefactored by the inverse relaxation timescales are linear.
Because the terms prefactored by $\gdot$ dominate the response of a
material to a fast shear startup and fast strain ramp, this
distinction will be important in what follows, particularly with
regards the onset of what we shall term `elastic instability'.

\subsection{General framework}
\label{sec: framework}

Motivated by the preceding discussion, we now outline a general
theoretical framework for the planar shear flow of complex fluids.
This will encompass as special cases the rolie-poly and Giesekus
models just described, as well as many other models for the rheology
of complex fluids. It is within this general framework that we shall
below perform a linear stability analysis to derive fluid-universal
criteria for the onset of shear banding in time-dependent
flows~\cite{Moorcroftetal2012}. Accordingly, the results that we
obtain should apply to all complex fluids that can be described by a
rheological constitutive equation of the highly general form that we
propose here.

We begin by combining all dynamical variables relevant to the fluid in
question into a state vector $\vecv{s}$. In a polymeric fluid this
will include all components of the viscoelastic conformation tensor
$\visc$ discussed above, $\vecv{s} = (\wxy,\wxx,\wyy,\ldots)^T$. In
soft glassy materials it would also include fluidity variables capable
of describing the slow evolution of a material into a progressively
more solid-like state~\cite{SuzanneInProgress}.

Next we define a projection vector $\vecv{p}=(1,0,0,\ldots)$ to select
out of this state vector the shear component $\wxy$ of the
viscoelastic conformation variable. The total shear stress
$\Sigma_{xy}=\Sigma$ is then written:
\beqn
\label{eqn: governing_eqn_force}
\Sigma(t) = G\vecv{p} \cdot \vecv{s}(y,t) + \eta \gdot(y,t).
\eeqn
Here and below we drop the $xy$ subscript from the shear component
$\Sigma$ of the total stress for clarity.

In a planar shear flow, the viscoelastic constitutive equation has the generalised form
\beqn
\partial_{t\,}\vecv{s} = \vecv{Q}(\vecv{s},\gdot) + D\partial_y^2\vecv{s}.
\label{eqn: governing_eqn_diffusive}
\eeqn
The choice of constitutive model then specifies the functional form of
$\vecv{Q}$. Depending on this choice we can obtain, for example, the
rolie-poly~\cite{Grahametal2003a}, Giesekus~\cite{Giesekus1982},
Johnson-Segalman~\cite{JS} or (for an infinite dimensional $\vecv{s}$)
soft glassy rheology model~\cite{SGR} and many more besides. The
compact notation introduced in this subsection therefore includes the
behaviour in shear of any fluid described in subsections A - D above.
Crucially, however, we shall not need to specify $\vecv{Q}$ in order
to perform our linear stability analysis for the onset of banding. Our
stability results will therefore be generic to all models described by
a constitutive equation of this highly general form.

\subsection{Units and parameters}

Throughout we choose units in which the rheometer gap width $L = 1$;
the elastic modulus $G=1$; and the terminal viscoelastic relaxation
time $\taud = 1$ (RP model) or $\lambda = 1$ (Giesekus model).

In our nonlinear simulations we shall set the value of the diffusion
constant $D$ such that the interface between the bands has a typical
lengthscale $\ell = 10^{-2}L$. In our linear stability analysis we
neglect the diffusive terms since they do not affect the results for
the long wavelength modes of interest here.

This leaves as parameters to be explored the solvent viscosity $\eta$,
the CCR parameter $\beta$ (nRP and sRP models), the stretch relaxation
time $\taur$ (sRP model) and the anisotropy parameter $\alpha$
(Giesekus model).

\section{Linear stability analysis}
\label{section: stability_analysis}

In this section we outline a linear stability analysis to determine
whether a state of initially homogeneous shear flow, which we shall
call the underlying ``base state'', becomes unstable to the growth of
heterogeneous perturbations that are the precursor of a shear banded
state. Distinct from more conventional linear stability analyses, we
are concerned here with a base state that is time-dependent,
comprising the initially homogeneous dynamical response of the fluid
following the imposition of a step stress, step strain, or shear
startup.  Accordingly, our analysis follows previous time-dependent
ones in Refs.~\cite{Fieldingetal2003a,Adamsetal2011,Manningetal2007a}.

\subsection{Equations of motion for heterogeneous perturbations}

Working within the general framework set out in Sec.~\ref{sec:
  framework} above, we start by expressing the response of the system
to the imposed flow protocol (step stress, strain ramp, shear startup)
as the sum of a time-dependent homogeneous base state plus any
(initially) small heterogeneous perturbations:
\beqna
\Sigma(t) &=& \base{\Sigma}(t),\nonumber\\
\gdot(y,t)&=& \base{\gdot}(t) + \sum_{n=1}^\infty \delta \gdot_n(t) \cos(n\pi y/L),\nonumber\\
\vecv{s}(y,t) &=& \vecv{\base{s}}(t) + \sum_{n=1}^\infty \delta\vecv{s}_n(t) \cos(n\pi y/L).
\label{eqn: LSA}
\eeqna
The first of these equations lacks the heterogeneous perturbations
seen in the other two because the total shear stress must remain
uniform across the sample, according to the force balance condition.
The time-dependence of the base state is clearly such that in a step
stress protocol $\base{\Sigma}(t)=\Sigma_0={\rm const.}$; during a
strain ramp $\base{\gdot}(t)=\gdot_0={\rm const.}$ with $\gdot_0=0$
post-ramp; and during a shear startup $\base{\gdot}(t)=\gdot_0={\rm
  const}$.

Substituting Eqns.~\ref{eqn: LSA} into Eqns.~\ref{eqn:
  governing_eqn_force} and~\ref{eqn: governing_eqn_diffusive},
neglecting the diffusive terms as noted above, and expanding in
successive powers of the magnitude of the small perturbations
$\delta{\gdot_n},\vecv{\delta s_n}$, we find at zeroth order that the
homogeneous base state obeys:
\beqna
\label{eqn: base}
\base{\Sigma}(t) &=& G\vecv{p} \cdot \base{\vecv{s}}(t) + \eta \base{\gdot}(t),\nonumber\\
\dot{\base{\vecv{s}}} &=& \vecv{Q}(\base{\vecv{s}},\base{\gdot}).
\eeqna
At first order, the heterogeneous perturbations obey
\beqna
\label{eqn: perturbation}
0&=&G\tens{p}\cdot \delta\vecv{s}_n(t)+\eta\delta\gdot_n(t),\nonumber\\
\dot{\delta\vecv{s}}_n &=& \tens{M}(t) \cdot \delta\vecv{s}_n + \tens{q}\delta{\gdot}_n,
\eeqna
in which  $\tens{M} =
\partial_{\vecv{s}\,}\vecv{Q}|_{\vecv{\base{s}},\base{\gdot}}$ and
$\vecv{q} = \partial_{\gdot}\vecv{Q}|_{\vecv{\base{s}},\base{\gdot}}$.
These two linearised equations can be combined to give
\beqn
\label{eqn: one}
\dot{\delta\vecv{s}}_n = \tens{P}(t) \cdot \delta\vecv{s}_n,
\eeqn
in which 
\beqn
\tens{P}(t) = \tens{M}(t) - \frac{G}{\eta}\vecv{q}(t)\, \vecv{p}.
\label{eqn: two}
\eeqn
We neglect terms of second order and above. 

In what follows our first objective will be to determine whether at
any time $t$ the heterogeneous perturbations
$\delta{\gdot}_n,\delta\vecv{s}_n(t)$ have a negative or positive rate
of growth, respectively indicating linear stability or instability to
the onset of shear banding. Our second objective is to relate the
onset of growth in these heterogeneous perturbations -- {\it i.e.},
the onset of shear banding -- to any distinctive signature in the
shape of the experimentally measured rheological response function as
specified by the evolution of the underlying homogeneous base state in
any given protocol.

We shall tackle these objectives using three different methods that we
cross-check against each other.  First, we denote by $\omega(t)$ the
real part the eigenvalue of $\tens{P}(t)$ that has the largest real
part at any time $t$. A positive $\omega(t)$ strongly suggests that
heterogeneous perturbations will be instantaneously growing at that
time $t$. This concept of a time-dependent eigenvalue must, however,
be treated with caution~\cite{Schmid2007a}.  Therefore second, and
better, we directly integrate the linearised Eqns.~\ref{eqn: one}
using an Euler time-stepping method, carefully converged with respect
to reducing the timestep. We examine the time-evolution of the
heterogeneous perturbations thereby calculated, to see whether at any
instant they are growing or decaying. Finally, we integrate the full
non-linear spatio-temporal Eqns.~\ref{eqn: governing_eqn_force}
and~\ref{eqn: governing_eqn_diffusive} using a Crank Nicolson
algorithm \cite{NumRecipes}, carefully checked for convergence with
respect to the size of the time- and space-steps.  The heterogeneous
part of the solution of this third method must coincide with the
results of the second method as long as the system remains in the
linear regime of small perturbations.

As written above, the stability matrix $\tens{P}$ appears to show no
dependence on the spatial lengthscale of the perturbation, as denoted
by $n$.  The same comment therefore applies to the eigenvalues of
$\tens{P}$.  This follows from our having neglected the diffusive term
in the viscoelastic constitutive equation before performing the
linearisation. Reinstating the diffusive term would simply transform
any eigenvalue $\omega \to \omega_n=\omega - D n^2\pi^2/L^2$ and act
to damp out any perturbations with a wavelength of order the
microscopic lengthscale $l$ or below. Accordingly the results of our
stability analysis apply only to perturbations of macroscopic
lengthscale, which are the ones of interest in the initial formation
of shear bands.

\subsection{Seeding the heterogeneous perturbations}

So far, we have discussed how to determine whether heterogeneous
perturbations to the homogeneous base state grow or decay over time.
We now consider how such perturbations are seeded into the system in
the first place. We identify several different possible physical
mechanisms for this, including (a) residual heterogeneities that
remain in the fluid following initial sample preparation, (b)
imperfect rheometer feedback or plate alignment, (c) true thermal
noise, or (d) slight rheometer device curvature in a curved Couette or
cone-and-plate geometry, which adds a small systematic perturbation to
the componentwise equations that we wrote above within the assumption
of a theoretically idealised planar geometry.

Of these, we model (a) by adding a small heterogeneous perturbation
once only, before the onset of deformation, by initialising
$\delta\vecv{s}_n(t=0)=q\vecv{X}\delta_{n1}$ for the Fourier modes of
the linearised equations, or correspondingly $\delta\vecv{s}(y,t=0) =
q\vecv{X}\cos(n\pi y/L)$ with $n=1$ in the full spatio-temporal
equations.  ($\delta_{nm}$ is the Kronecker delta function, equal to
$1$ if $n=m$ and equal to $0$ otherwise.)  This initial perturbation
has magnitude $q$, which we treat as a parameter of our study.
$\vecv{X}$ is a vector of the same dimension as the vector $\tens{s}$.
Each of its components is a random number drawn from a uniform
distribution of mean 0 and width 1.  We choose to seed only the lowest
mode $n=1$ because it is always the most unstable (or least stable)
one: as discussed above, the diffusive terms render modes of higher
$n$ less unstable (or more stable).

The results presented below follow the method of seeding (a) just
described unless otherwise stated. In some cases we also check these
results against scenarios (b) and (c), modelled by adding a small
heterogeneous perturbation $q\sqrt{dt}\vecv{X}\delta_{n1}$ at every
timestep (of duration $dt$) to the Fourier modes of the linearised
equations, or correspondingly $q\sqrt{dt}\vecv{X}\cos(\pi y/L)$ at
every timestep in the full spatio-temporal simulation. A new random
vector $\vecv{X}$ is selected at each timestep. In this case we first
evolve the system under conditions of no applied flow or loading but
subject to this continuous noise, until a statistically steady state
is reached that correctly captures the fluctuation spectrum of the
system in zero shear. We then evolve the chosen flow protocol, also
subject to this continuous noise.

For the linearised equations subject to continuous noise, it is
furthermore possible to perform an upfront analytical average (denoted
$\langle \rangle$) over infinitely many noise histories by evolving
the variance of the perturbations:
\beqn
\partial_{t\,}\tens{S} = \tens{P} \cdot \tens{S} + \tens{S} \cdot \tens{P}^{T} + \tens{N},
\label{eqn: S}
\eeqn
in which $\tens{S}(t) = \langle \vecv{\delta s_n}(t) \cdot
\vecv{\delta s_n}^T(t) \rangle$ and $\tens{N}(t)$ is a diagonal matrix
characterising the amplitude of the added noise.  Using the linearised
force balance condition we then easily obtain the evolution of the
variance of the shear rate perturbations $\langle \delta \gdot_{n}^2
\rangle(t)$.

\subsection{Reporting the heterogeneous perturbations}

For the linearised system subject to seeding (a) above we report the
size of the shear rate heterogeneity as $| \delta \gdot_{n=1}|(t)$.
For a linearised system subject to continuous noise (b, c) we report
$\sqrt{\langle \delta \gdot_{n=1}^2 \rangle}(t)$. In the full
nonlinear spatiotemporal simulation we quantify the degree of
heterogeneity by the difference at any time between the maximum and
minimum values of the shear rate across the cell:
\beqn
\Delta_{\gdot}(t) = \gdot_{\rm max} - \gdot_{\rm min}.
\label{eqn: dob}
\eeqn
We often below refer to this quantity as the `degree of banding'. 

We have checked that as long as the nonlinear simulation remains in
the linear regime of small heterogeneity, the $\Delta_{\gdot}$ that it
predicts evolves in the same way as the perturbations calculated in
the corresponding linearised calculation, up to a constant prefactor
$O(1)$.

\section{Results: step stress}
\label{section: stepstress}

In this section we present our results for time-dependent shear
banding during a system's creep response following the imposition of a
step shear stress to a previously unloaded sample,
$\Sigma(t)=\Sigma_0\Theta(t)$.  We start in Sec.\ref{section:
  criterion_step_stress} by developing an analytical criterion for the
onset of banding. In Secs.~\ref{section: RP_stepstress}
and~\ref{section: giesekus_stepstress} we give numerical results to
support this prediction, in the rolie-poly and Giesekus models
respectively.

\subsection{Criterion for shear banding following a step stress}

\label{section: criterion_step_stress}

Here we develop a criterion for the onset of shear banding in the step
stress protocol. We do so by considering an underlying base state of
initially homogeneous flow response to the applied loading, and the
dynamics of small heterogeneous perturbations about this base state.

Following the imposition of a step shear stress to a previously
unloaded sample, it is easy to show by time-differentiating
Eqns.~\ref{eqn: base} subject to the constraint
$\base{\Sigma}(t)=\Sigma_0$ that any underlying base state of
initially homogeneous flow response must obey
\beqn
\frac{d}{dt} \vecv{\base{\dot{s}}} = \left(\tens{M} - G\vecv{q} \vecv{p}/\eta\right) \cdot \vecv{\base{\dot{s}}}.
\eeqn
Eqns.~\ref{eqn: one} and~\ref{eqn: two} together show that any
heterogeneous perturbations to this base state must obey
\beqn 
\frac{d}{dt} \delta\vecv{s}_n = \left(\tens{M} - G\vecv{q} \vecv{p}/\eta\right) \cdot \delta\vecv{s}_n.
\label{eqn: creep_lin}
\eeqn

Comparing these two equations, we see that the heterogeneous
perturbations $\delta\vecv{s}_n$ obey the same dynamical equation as
the time-derivative of the homogeneous base state
$\vecv{\base{\dot{s}}}$.  Combined with the force balance condition
(Eqn.~\ref{eqn: force_balance}) and its linearised counterpart, this
means that heterogeneous shear rate perturbations must grow, and shear
bands develop,
\beqn
\frac{d\delta\gdot_n}{dt}/ \delta\gdot_n > 0,
\eeqn
in any regime where
\beqn
\frac{d^2 \base{\gdot}}{dt^2} / \frac{d\base{\gdot}}{dt} > 0.
\label{eqn: stepstress_criterion}
\eeqn

This important result tells us that a state of initially homogeneous
creep response to an imposed step stress becomes linearly unstable to
the onset of shear banding whenever its shear rate signal
$\base{\gdot}(t)$ is simultaneously upwardly curving and upwardly
sloping in time.  (Alternatively $\base{\gdot}(t)$ may be
simultaneously downwardly curving and downwardly sloping, though in
practice we have never seen this numerically.)

\iffigures
\begin{figure}[tbp]
\includegraphics[width=7cm,height=5cm]{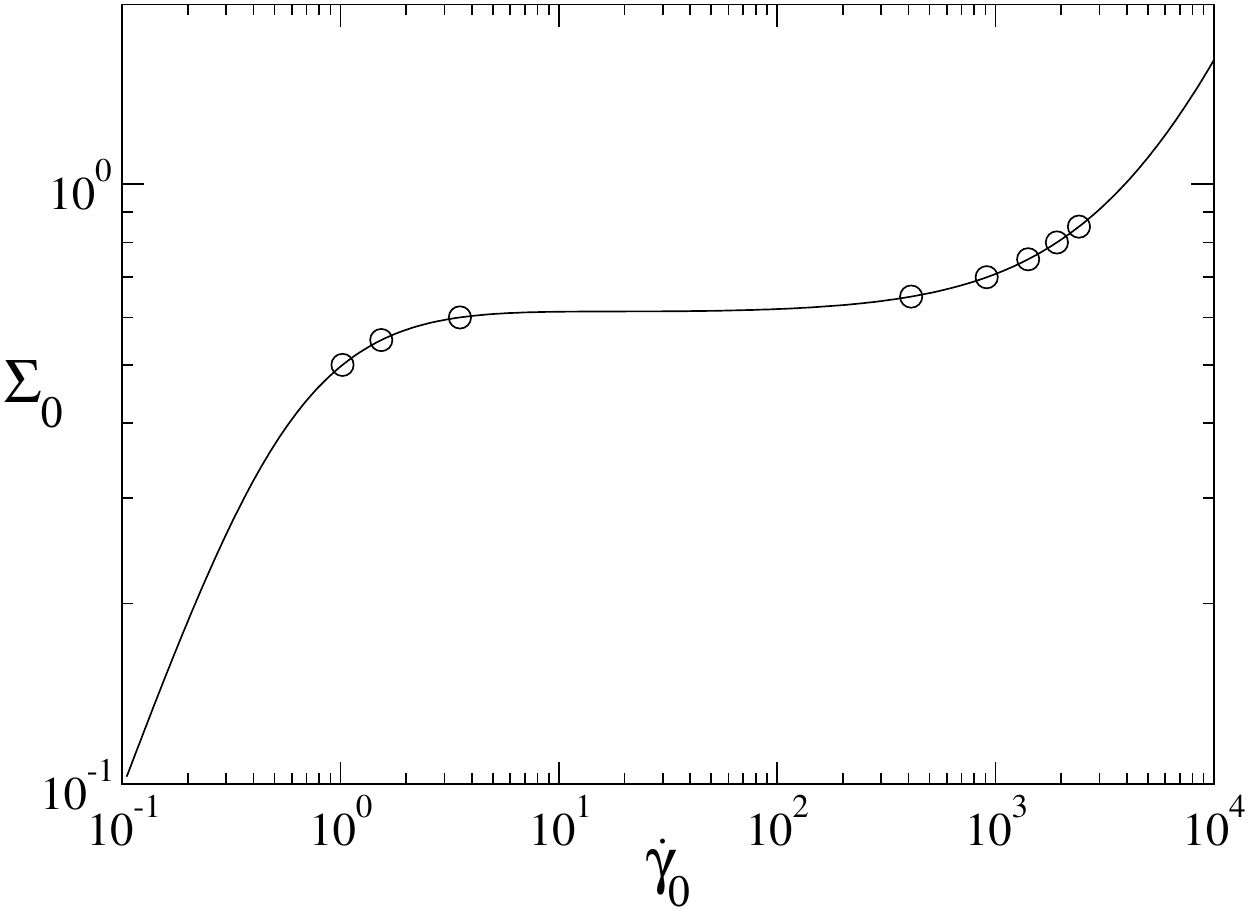}
\caption{Steady state constitutive curve in the rolie-poly model.
  Symbols correspond to steady states from Fig.~\ref{fig: RP1a}.
  Parameters: $\beta = 0.8$, $\eta = 10^{-4}$, $\taur = 0.0$.}
 \label{fig: RP1b}
\end{figure}
\fi

\iffigures
\begin{figure}[tbp]
\includegraphics[width=7cm,height=5cm]{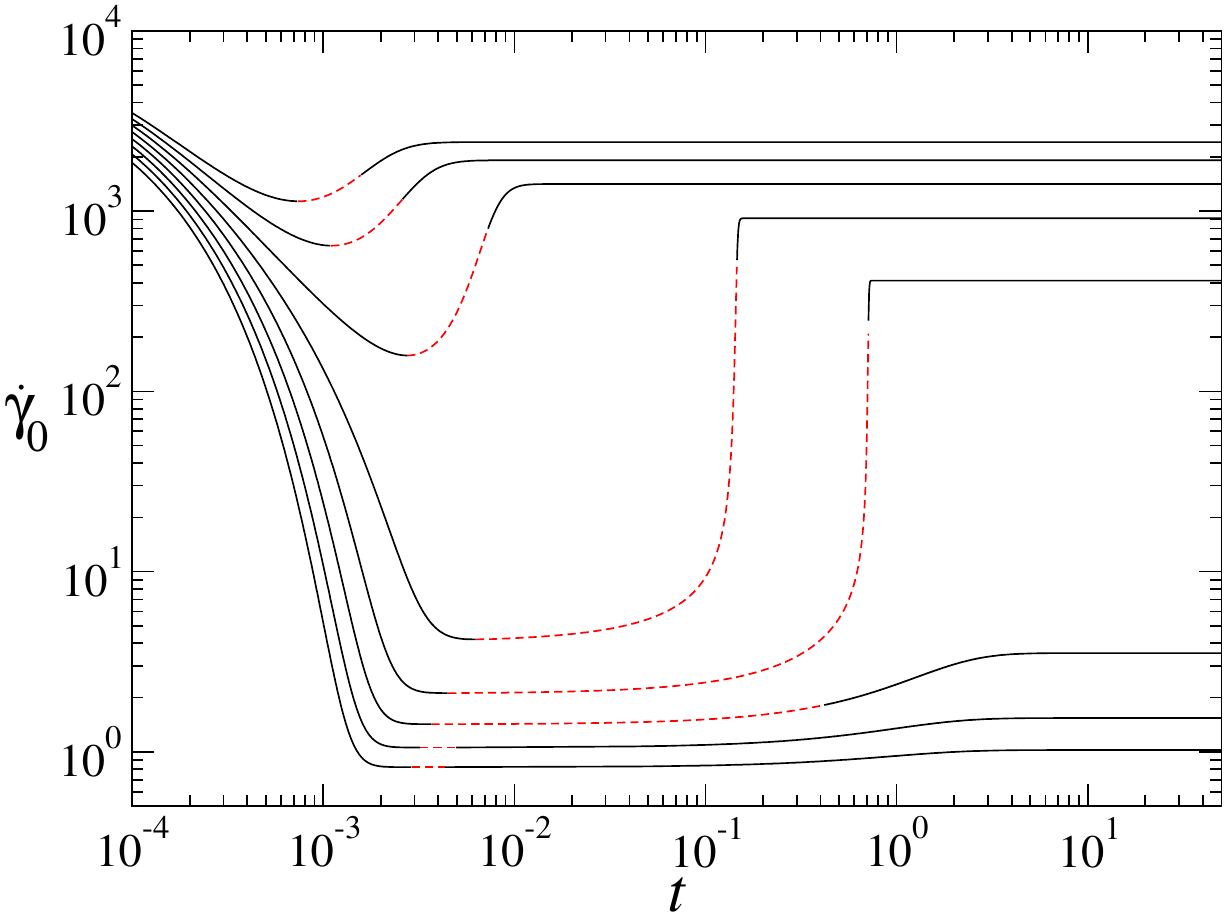}
\caption{Shear rate as a function of time for imposed shear stresses
  $\Sigma_0 = 0.5,0.55,...,0.85$ (curves bottom to top at fixed $t$) in
  the (homogeneously constrained) RP model. Dashed lines show regions
  of linear instability, $\partial_{t}^{2} \gdot_0 / \partial_{t} \gdot_0
  > 0$. Steady states correspond to the circles in Fig.~\ref{fig: RP1b}.
  Parameters: $\beta = 0.8$, $\eta = 10^{-4}$, $\taur = 0.0$.}
 \label{fig: RP1a}
\end{figure}
\fi

\iffigures
\begin{figure*}[tbp]
  \includegraphics[width=14.0cm, height=10.0cm]{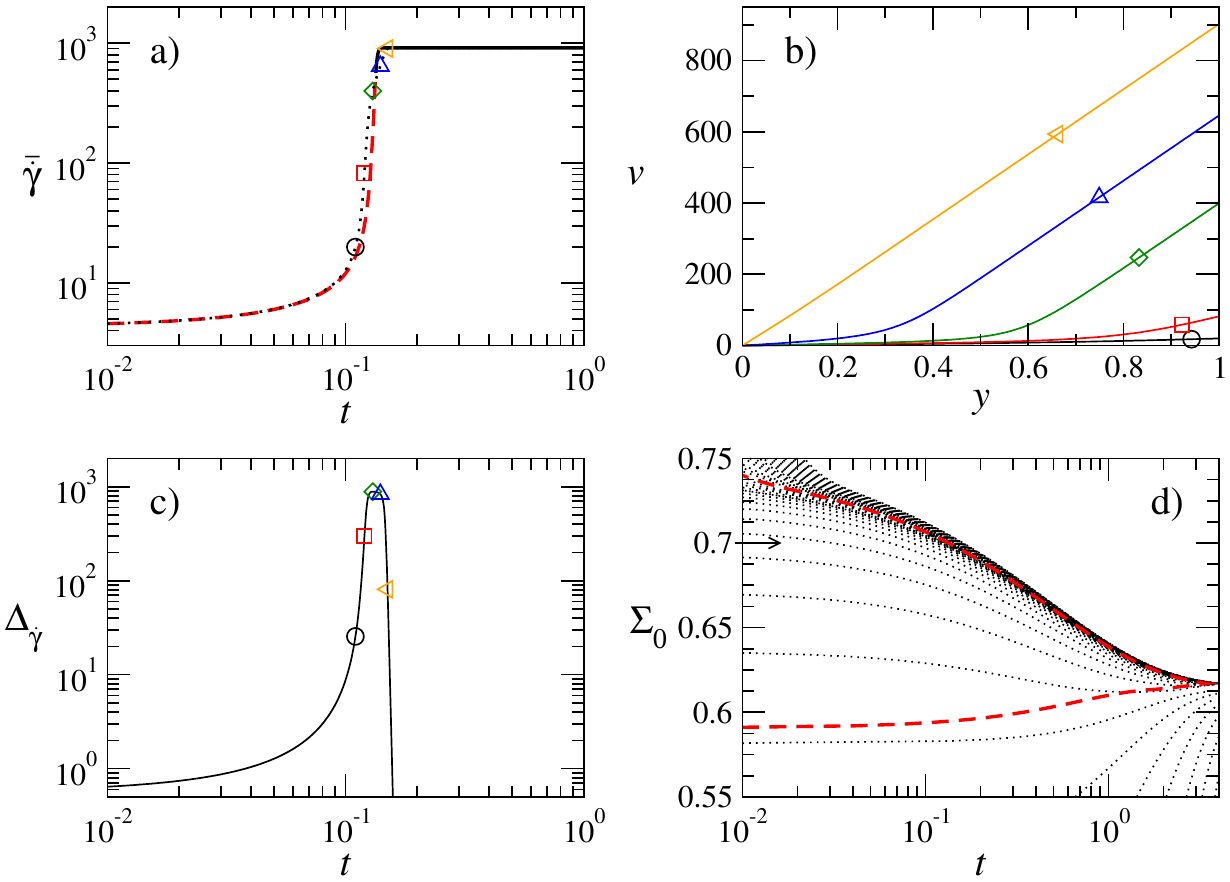}
  \caption{Step stress of amplitude $\Sigma_0=0.7$ in the RP model: (a) Thick line: global shear rate
    in a homogeneously constrained system, with the dashed region denoting $\partial_{t}^{2}\gdot_0 /
    \partial_{t}\gdot_0 > 0$. Dotted line shows corresponding signal
    with heterogeneity allowed, $\gdotbar$. (b) Snapshots of the
    velocity profile at times corresponding to symbols in (a).  (c)
    Corresponding degree of banding: $\Delta_{\gdot} = \gdot_{\rm max}
    - \gdot_{\rm min}$. (d) Banding dynamics in the full plane of
    stress versus time. A horizontal slice across this plane
    corresponds to a single run at a constant imposed stress
    $\Sigma_0$, in which we integrate the linearised equations for the
    dynamics of the heterogeneous perturbations.  Dotted lines are
    contours of equal $|\delta\gdot|_{n=1}(t) =
    |\delta\gdot|_{n=1}(0)2^M$ for integer $M$. (We show
    only contours $M>-50$, thereby cutting off the final stage of the
    decay at the right hand side of the graph.) The thick dashed line
    shows where the base state $\partial_t^{\,2} \gdot_0/\partial_t
    \gdot_0 = 0$, with linear instability to the left of it. The arrow
    denotes the stress value explored in detail in subfigures (a) to
    (c).  Parameters: $\beta = 0.8$, $\eta = 10^{-4}$, $\taur = 0.0$,
    $q = 0.1$.  }
  \label{fig: RP2}
\end{figure*}
%
\fi

How does this time-differentiated creep curve $\base{\gdot}(t)$ of the
underlying homogeneous base state relate to the time-differentiated
creep curve $\gdotbar(t)$ measured experimentally by recording the
motion of the rheometer plates?  Clearly, in any regime before banding
sets in these two quantities coincide by definition. Once shear
banding fluctuations have grown appreciably into the nonlinear regime,
however, the two need not coincide. Nevertheless, in our numerical
studies of step stress in the rolie-poly and Giesekus models (and also
of the soft glassy rheology model reported
elsewhere~\cite{SuzanneInProgress}) we have never observed the
globally measured bulk shear rate signal to be strongly affected, in
overall shape at least, by the presence of shear banding within the
fluid.  Accordingly we can take Eqn.~\ref{eqn: stepstress_criterion}
to apply also to the experimentally measured signal $\gdotbar(t)$.
Experimentalists should therefore be alert to the onset of shear
banding in any creep experiment where the time-differential of the
measured creep curve simultaneously shows upward slope and upward
curvature: bulk rheological data can be used as a predictor of the
presence of shear banding, even in the absence of accompanying
velocimetric data.

\subsection{Numerical results: rolie-poly model}
\label{section: RP_stepstress}

Having developed an analytical criterion for the onset of shear
banding following the imposition of a step stress, we now present
numerical results that support it.  Fig.~\ref{fig: RP1b} shows the
underlying constitutive curve of stress as a function of strain rate
in the rolie-poly model for a value of the CCR parameter $\beta=0.8$.
Because this curve is monotonic, the eventual steadily flowing state
is homogeneous.

Nonetheless for imposed stress values in the relatively flat region of
this curve, we might expect shear bands to form transiently as the
system evolves towards its steady state on this ultimate constitutive
curve.  Motivated by this expectation, we now study numerically the
step stress protocol for the stress values denoted by circles in
Fig.~\ref{fig: RP1b}.

We report first results for the underlying time-dependent base state
of homogeneous creep response to this imposed load, obtained in a
numerical calculation in which the flow is artificially constrained to
remain homogeneous. The shear rate evolution $\base{\gdot}(t)$ in this
case is shown in Fig.~\ref{fig: RP1a}. Immediately after loading the
solvent bears all the applied stress and $\base{\gdot}(t=0^+)=
\Sigma_0/\eta$. The shear rate then shows a rapid early decay on a
timescale that appears numerically to scale as
$O(\sqrt{\eta\,\tau_d/G})$, but that may not be accessible
experimentally due to inertial effects such as creep ringing that we
have neglected here.  (Our numerics set any inertial timescales to
zero.) Following this fast initial drop, the shear rate subsequently
displays a regime of simultaneous upwards curvature and upwards
slope, shown by dashed lines in the figure. In this regime, this
underlying base state of homogeneous flow is predicted by our
criterion~(\ref{eqn: stepstress_criterion}) above to become linearly
unstable to the onset of shear rate heterogeneity.

Accordingly, in Figure \ref{fig: RP2}a-c) we show the results of a
fully nonlinear simulation that now permits heterogeneity in the
flow-gradient direction. Subfigure (a) shows the evolution of the
global shear rate $\gdotbar(t)$ for a single stress value
$\Sigma=0.7$, near the point of weakest slope of the constitutive
curve in Fig.~\ref{fig: RP1a}. As can be seen, this bulk rheological
signal differs little from that given by our earlier homogeneous
calculation: $\gdotbar(t)\approx\base{\gdot}(t)$, even in the regime
where bands form and the two signals might be expected to differ. This
supports our claim made above that the theoretical criterion of
Eqn.~\ref{eqn: stepstress_criterion}, which strictly applies only to
the underlying base state $\base{\gdot}(t)$, can also be applied to
the experimentally measured bulk signal $\gdotbar(t)$.

Subfigure (b) shows snapshots of the velocity profiles that accompany
the bulk signal of (a). These clearly exhibit macroscopic shear
banding. Plotting the associated degree of banding $\Delta_{\gdot}(t)
= \gdot_{\rm max} - \gdot_{\rm min} $ as a function of time in (c), we
find good agreement with our prediction (\ref{eqn:
  stepstress_criterion}). Banding sets in once $\gdotbar$ shows upward
curvature. The flow then returns to be homogeneous once $\gdotbar$
exhibits downward curvature during the final stage of its evolution to
steady state.

Subfigure~\ref{fig: RP2}d) summarises the shear banding dynamics of
the system across a range of stress values, in the plane of stress
versus time.  Any horizontal slice across this plane represents a
single creep run at a constant value of the imposed stress $\Sigma$,
as discussed in subfigures (a)-(c).  The thick dashed line encloses to
its left the regime of linear instability to the onset of banding, in
which heterogeneous shear rate perturbations are predicted to grow.
This line is obtained by applying our criterion~(\ref{eqn:
  stepstress_criterion}) to the base state signal calculated
numerically in a series of runs at closely spaced values of the
imposed stress. To make an exploration of the dynamics of shear
banding perturbations feasible in this full plane of stress versus
time, we integrated the linearised equations of motion (\ref{eqn:
  creep_lin}).  (Performing the full nonlinear and spatially aware
simulation across a wide range of closely spaced stress values would
be much more time consuming computationally.) The dotted lines show
contours of equal $|\delta\gdot|_{n=1}(t) = |\delta\gdot|_{n=1}(0)2^M$
for integer $M$.  The region of growth in these
perturbations agrees well with our analytical criterion enclosed by
the dashed line.  It corresponds to stress values $0.61 \lesssim
\Sigma \lesssim 0.75$ around the region of weakest slope of the
constitutive curve.

\iffigures
\begin{figure}[tbp]
  \includegraphics[width = 7.0cm, height = 7.0cm]{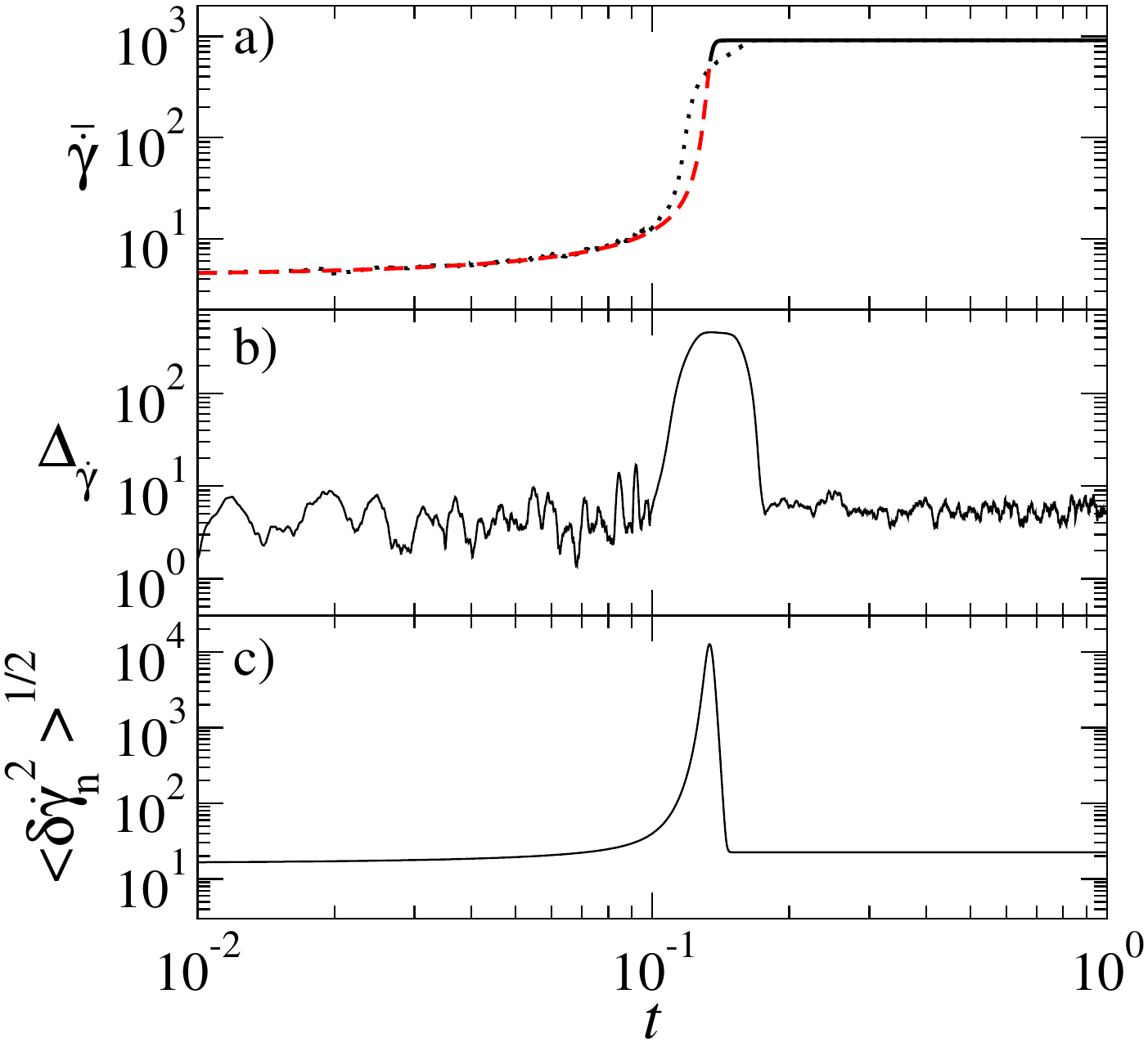}
  \caption{Step stress in the RP model with noise added at each timestep.
    (a) Thick line: global shear rate $\gdot_0$ in a homogeneously
    constrained system, with the dashed region denoting
    $\partial_{t}^{2}\gdot_0 /
    \partial_{t}\gdot_0 > 0$. Dotted line shows corresponding signal
    with heterogeneity allowed $\gdotbar$.  (b) Degree of shear
    banding: $\Delta_{\gdot} = \gdot_{\rm max} - \gdot_{\rm min}$ from
    a fully nonlinear simulation with $q = 0.1$. (Here a running
    average over data captured at frequent points in time is used,
    checked for qualitative convergence with respect to the capture
    frequency and running average range.) (c) Shear rate perturbation
    $\sqrt{\langle\delta\gdot_{n}^2 \rangle}$ of the linearised system
    found by integrating Eqn \ref{eqn: S}. Magnitude of noise $q =
    10^{-5}$.  Values of model parameters as in Fig.~\ref{fig: RP2}.}
  \label{fig: RP3}
\end{figure}
\fi

Time-dependent shear banding during a sharp increase in the shear rate
response $\gdotbar(t)$ following the imposition of a step stress has
been reported experimentally in polymeric systems in
Refs.~\cite{Wangetal2009a, Huetal2007a, Wangetal2003a, Huetal2008a,
  Wangetal2008c, Huetal2005a, Huetal2010a, Wangetal2009d}.

The results in Fig.~\ref{fig: RP2} apply to a system in which a
heterogeneous perturbation is seeded once only, at the initial time
$t=0$. In practice, such a situation might correspond to the sample
being left in a slightly heterogeneous state as a result of the
experimental protocol by which it is initially loaded into the
rheometer.  Alternatively, perturbations may be seeded continuously
during the experiment due to imperfect rheometer feedback. To model
this we also performed calculations in which small heterogeneous
perturbations are added at every timestep.  Pleasingly, we find
qualitatively similar results: compare Figs.~\ref{fig: RP2}
and~\ref{fig: RP3}.

\subsection{Numerical results: Giesekus model}
\label{section: giesekus_stepstress}

We now discuss our numerical results for an imposed step stress in the
Giesekus model. To ensure a fair comparison with our study of the
rolie-poly model just described, we use a value of the anisotropy
parameter $\alpha$ such that the underlying constitutive curve is
monotonic and as closely resembling that of Fig.~\ref{fig: RP1b} as
possible.  Also as before, we set our value for the imposed stress to
be in the region of weakest slope in this curve.

\iffigures
\begin{figure*}[tbp]
  \centering
  \includegraphics[width=12.0cm, height=6.5cm]{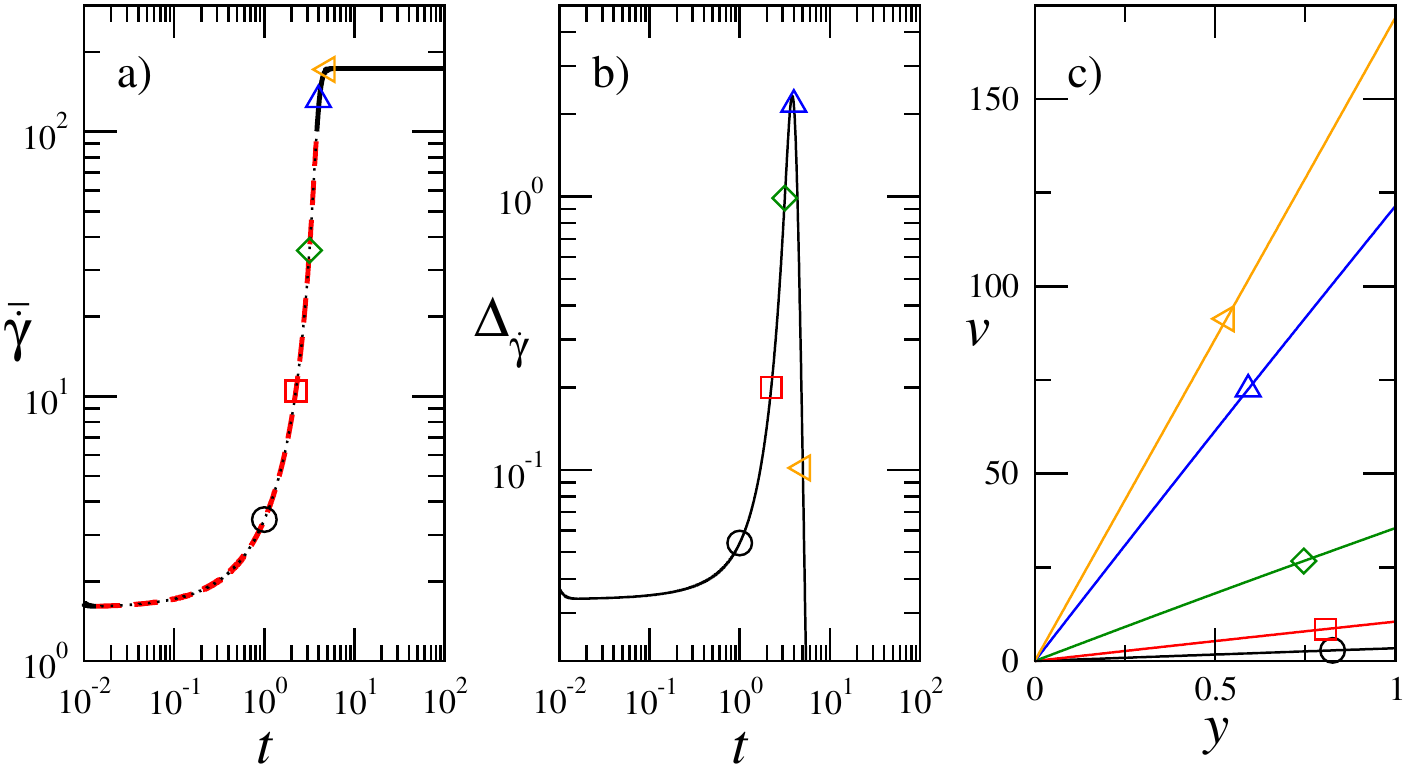}
  \caption{Step stress in the Giesekus model.
(a) Thick line:
    shear rate response in a homogeneously constrained system, with the dashed region denoting $\partial_{t}^{2}\gdot_0 /
    \partial_{t}\gdot_0 > 0$. Dotted line shows corresponding signal
    with heterogeneity allowed, which is now indistinguishable from
    the homogeneous signal.  (b) Degree of banding: $\Delta_{\gdot} =
    \gdot_{\rm max} - \gdot_{\rm min}$. (c) Snapshots of the velocity
    profile at times corresponding to symbols in (a) and (b).
    Parameters: $\alpha = 0.6$, $\eta = 10^{-3}$, $\Sigma = 1.0$,
    $q=0.1$.}
  \label{fig: Gies1}
\end{figure*}
\fi

The shape of the shear rate response to this imposed step stress is
qualitatively similar to that seen in the RP model, in particular in
showing a regime of upward curvature. Compare Figs.~\ref{fig: RP3}a)
and~\ref{fig: Gies1}a).  In principle this upward curvature renders a
state of initially homogeneous flow unstable to the development of
shear bands.  Indeed, shear rate heterogeneities do start to grow as a
result of this linear instability. However we have never found them to
grow sufficiently large as to give `significant' shear banding in our
numerical simulations of the Giesekus model.  The degree of banding
$\Delta_{\gdot}(t) = \gdot_{\rm max} - \gdot_{\rm min}$ never exceeds
$5$\% of the global shear rate $\gdotbar(t)$ averaged across the
sample, and thus would be hard to detect experimentally~\footnote{We
  note that, within the linear regime, the degree of banding
  $\Delta_{\gdot}$ scales linearly with the magnitude of the initial
  noise $q$.  Therefore, in order to make comparisons with the RP
  model the results presented in sections~\ref{section: RP_stepstress}
  and~\ref{section: giesekus_stepstress} section have $q = 0.1$. We
  note that $q$ values much larger than this are unrealistic for
  comparison to experiment.}.  Contrast the results for
$\Delta_{\gdot}(t)$ and $v(y)$ in Fig.~\ref{fig: Gies1} with their
counterpart for the RP model in Fig.~\ref{fig: RP2}c). By repeating
this numerical calculation across a wide range of values of
$\alpha,\eta$, and $\Sigma_0$, we checked that this conclusion of
negligible banding is general for this protocol in the Giesekus model.

The reason for this striking difference in shear banding behaviour
between the two models, despite their differentiated creep response
curves $\gdot_0(t)$ having the same upwardly curving shape, is that
the maximal value of the curvature in $\gdot(t)$ is always much
smaller in the Giesekus model than the RP model. Compare
Figs.~\ref{fig: RP2}a) and \ref{fig: Gies1}a).  Correspondingly, the
resulting maximum degree of shear banding is much smaller.

\iffigures
\begin{figure}[tbp]
  \centering
  \includegraphics[width=7.0cm, height=7.0cm]{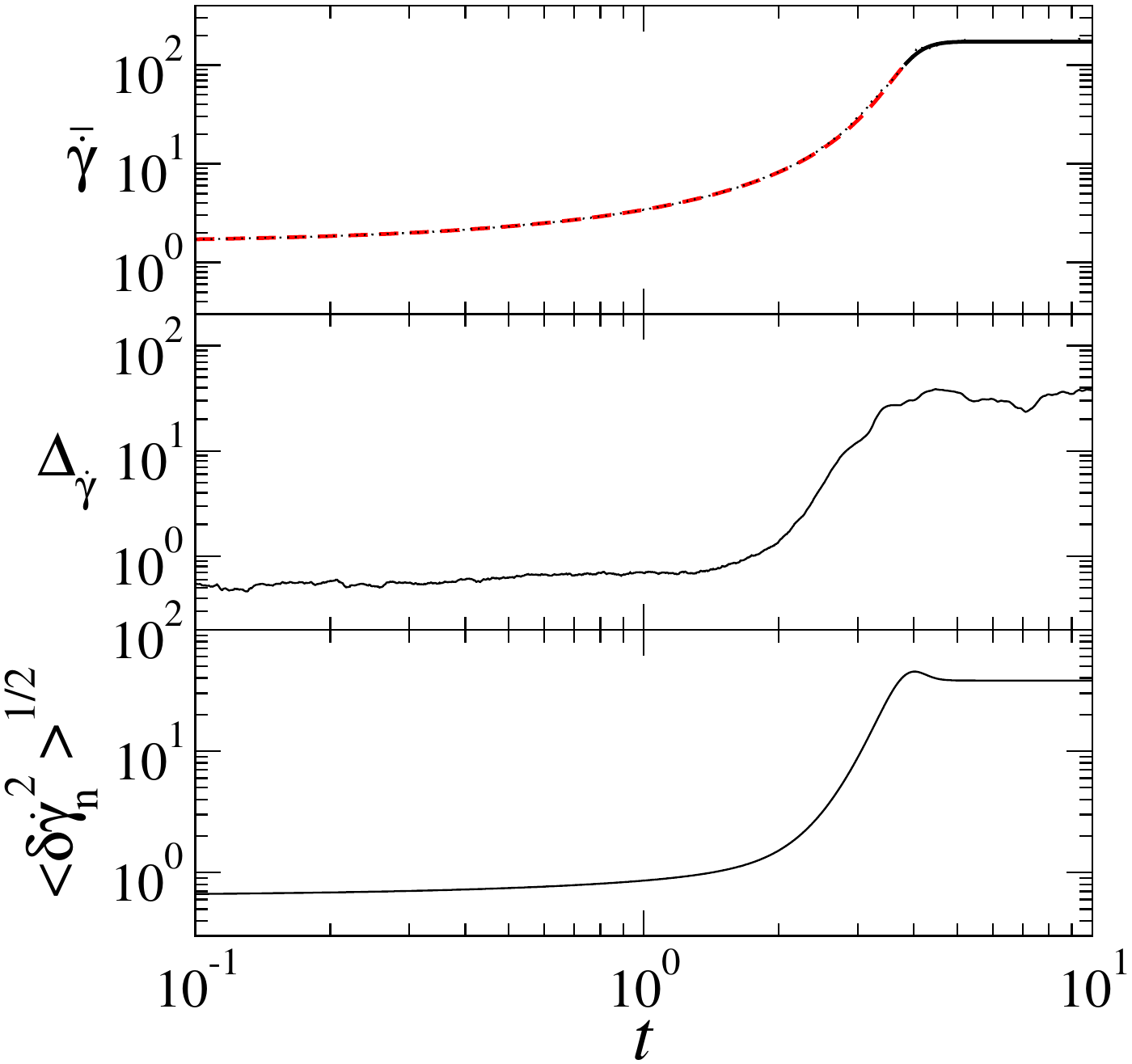}
  \caption{Step stress in the Giesekus model with parameters as in
    Figure \ref{fig: Gies1}. 
Top: Thick line shows the shear rate
    $\gdot_0$ in a homogeneously constrained system, with the dashed region denoting $\partial_{t}^{2}\gdot_0 /
    \partial_{t}\gdot_0 > 0$. Dotted line shows corresponding signal
    with heterogeneity allowed $\gdotbar$, now indistinguishable from
    the homogeneous signal.  Middle: degree of shear banding:
    $\Delta_{\gdot} = \gdot_{\rm max} - \gdot_{\rm min}$ from the
    nonlinear simulation with $q = 0.1$ (Here a running average over
    data captured at frequent points is used, checked for qualitative
    convergence with respect to the capture frequency and running
    average range.) Bottom: Shear rate perturbation $\sqrt{\langle
      \delta\gdot_{n=1}^2 \rangle}$ in linearised system found by
    integrating Eqns.~\ref{eqn: S}, $q = 10^{-5}$.  }
  \label{fig: Gies2}
\end{figure}
\fi

In Fig.~\ref{fig: Gies1}, heterogeneous perturbations are seeded once
only, at the initial time $t=0$.  By comparison in Fig.~\ref{fig:
  Gies2} the system is first evolved to steady state at $\Sigma=0$
with small perturbations added at every timestep, before being evolved
at the chosen $\Sigma=\Sigma_0$ with perturbations again added
continuously at every timestep. Although we never found `significant
banding' ($>5\%$) in the Giesekus model with this noise history
either, there is nonetheless an interesting feature not seen in the RP
model. In particular, the amplitude of heterogeneity in steady flow is
much greater than in the unsheared system at rest.  (In contrast, in
the RP model the magnitude of heterogeneity in steady shear is
comparable to that in zero shear, recall Fig.~\ref{fig: RP3}.)
Whether or not these fluctuations would be large enough to be detected
by sensitive velocimetry, this remains an interesting feature of the
Giesekus model.

\section{Results: strain ramp}
\label{section: rampstrain}

In this section we consider a strain ramp protocol in which a
previously undeformed sample is subject to an applied shear at rate
$\gdot_0$ by moving the top plate at speed $\gdot_0L$ for times
$0<t<t^*$. After this the plate is held fixed, giving a total applied
strain amplitude $\gamma^*=\gdot_0 t^*$. The limit $\gdot_0\to\infty$,
$t^*\to 0$ at fixed $\gamma^*$ gives a theoretically idealised step
strain. We focus here on ramp rates that are finite but nonetheless
always fast compared to the terminal relaxation time. Therefore we
impose $\gdot_0\taud\gg 1$ in the RP model and $\gdot_0\lambda \gg 1$
in the Giesekus model. (For the RP model, we then separately
distinguish between ramps for which $\gdot_0\taur \ll 1$ and
$\gdot_0\taur\gg 1$.) Following such fast ramps, we develop a
criterion for the transient appearance of shear bands as the system
relaxes back to equilibrium post-ramp.

\subsection{Criterion for instability after a fast strain ramp}
\label{section: criterion_stepstrain}

We start by writing our general governing Eqns.~\ref{eqn:
  governing_eqn_force} and~\ref{eqn: governing_eqn_diffusive} in a
form that emphasizes the additive loading and relaxation dynamics that
obtain in all constitutive equations of which we are aware:
\beqna
\Sigma(t) &=& G\vecv{p} \cdot \vecv{s}(y,t) + \eta \gdot(y,t),
\label{eqn: FB_rewrite}\\
\partial_{t\,}\vecv{s} &=& \gdot \vecv{S} (\vecv{s}) - \frac{1}{\tau} \vecv{R}(\vecv{s})\label{eqn: governing_eqn_rewrite}.
\eeqna
Here $\tau$ is the terminal relaxation time. (Two relaxation times, as
in the sRP model, are included in this notation by writing
$\vecv{R}=\vecv{R}_1+\tfrac{\tau}{\taur}\vecv{R}_2$.) For the purposes
of the linear stability calculation that follows we have neglected
small diffusive terms in Eqn.~\ref{eqn: governing_eqn_rewrite}, as
discussed above.

To develop our criterion for the onset of shear banding we follow our
usual linear stability procedure of considering an underlying base
state of initially homogeneous flow response to the imposed
deformation, and the dynamics of heterogeneous perturbations to this
base state that might grow into observable shear banding. 

Accordingly we write
\beqna
\Sigma(t) &=& \base{\Sigma}(t),\nonumber\\
\gdot(y,t)&=& \base{\gdot}(t) + \sum_{n=1}^\infty \delta \gdot_n(t) \cos(n\pi y/L),\nonumber\\
\vecv{s}(y,t) &=& \vecv{\base{s}}(t) + \sum_{n=1}^\infty \delta\vecv{s}_n(t) \cos(n\pi y/L).
\label{eqn: LSAr}
\eeqna
We then substitute these into the governing equations (\ref{eqn:
  FB_rewrite}) and (\ref{eqn: governing_eqn_rewrite}), and expand in
successive powers of the amplitude of the perturbations.

The zeroth order equations in this expansion govern the evolution of
the base state.  During the ramp this evolves according to
\beqna
\frac{d\vecv{\base{s}}}{d \base{\gamma}} &=& \vecv{S} (\vecv{\base{s}}) - \frac{1}{\tau\base{\gdot}} \vecv{R}(\vecv{\base{s}})\nonumber\\
&\approx& \vecv{S} (\vecv{\base{s}}).
\label{eqn: step_base}
\eeqna
In the second line we have specialised to the fast ramps of interest
here, for which the loading dynamics dominates.
Denoting the base state immediately as the ramp ends
$\base{\vecv{s}}(t=t^{*-})=\base{\vecv{s}}^*$, the dynamics of the base
state immediately prior to the end of the ramp obeys
\beqn
\frac{d\vecv{\base{s}}}{d\base{\gamma}}\mid_{t^{*-}} = \vecv{S}(\vecv{\base{s}^*}).
\label{eqn: the_first}
\eeqn
Post-ramp the base state relaxes back to equilibrium according to
\beqna
\frac{d\vecv{\base{s}}}{dt} = -\frac{1}{\tau}\vecv{R}(\vecv{\base{s}}).
\label{eqn: after}
\eeqna

Having discussed the evolution of the underlying homogeneous base
state we now turn to the linearised dynamics of the heterogeneous
perturbations. These are specified by the first order equations in the
amplitude expansion just discussed. (Recall Eqns.~\ref{eqn: one}
and~\ref{eqn: two}.)  Post ramp, these perturbations obey
\beqna
\frac{d\vecv{\delta s_n}}{dt} &=& \left[-\frac{G}{\eta}\,\vecv{S}(\vecv{\base{s}})\, \vecv{p} - \frac{1}{\tau}\partial_{\vecv{s}}\,\vecv{R}\mid_{\vecv{\base{s}}}\right] \cdot \vecv{\delta s_n} \, \nonumber \\
 &\simeq& - \frac{G}{\eta}\, \vecv{S}(\vecv{\base{s}}) \, \vecv{p} \cdot  \vecv{\delta s_n}.
\eeqna
The approximation on the second line is valid for small values of the
Newtonian viscosity compared with the zero shear polymer viscosity,
$\eta \ll G\tau$, which is a good approximation in most complex
fluids.  Because the base state $\base{\vecv{s}}$ is continuous across
the end of the ramp, it follows that {\em immediately} post-ramp the
perturbations obey
\beqn
\frac{d\vecv{\delta s_n}}{dt}\mid_{t^{*+}} = -\frac{G}{\eta}\vecv{S}(\vecv{\base{s}}^*) \vecv{p} \cdot \vecv{\delta s_n}.
\label{eqn: the_second}
\eeqn

Combining Eqn.~\ref{eqn: the_first} for the dynamics of the base state
immediately before the ramp ends with Eqn.~\ref{eqn: the_second} for
the dynamics of the heterogeneous perturbations immediately post-ramp,
we get
\beqn
\frac{\vecv{\delta s_n}}{dt}\mid_{t^{*+}} =
  -\frac{G}{\eta}\frac{d\vecv{\base{s}}}{d\base{\gamma}}\mid_{t^{*-}} \vecv{p} \cdot
  \vecv{\delta s_n}
\eeqn
Projecting out the first component of this equation using the operator
$\vecv{p}$, and appealing to the linearity of the force balance
condition (\ref{eqn: FB_rewrite}), it is easy to show finally that
shear rate perturbations obey, immediately post-ramp:
\beqn
\frac{\deltag}{dt}\mid_{t^{*+}} = -\frac{1}{\eta} \frac{\partial \base{\Sigma}}{\partial\base{\gamma}}\mid_{t^{*-}} \deltag.
\label{eqn: step_strain_criterion_growth}
\eeqn

This important result tells us that, immediately after a strain ramp
has ended, an initially homogeneous flow state will be linearly
unstable to the onset of shear banding if the shear stress had been
decreasing in strain immediately before the ramp ended:
\beqn
\frac{\partial \base{\Sigma}}{\partial \base{\gamma}}\mid_{t^{*-}}  < 0.
\label{eqn: step_strain_criterion}
\eeqn
This is consistent with the original insight of Marrucci and Grizzuti
\cite{MarrucciGrizzuti1983a} in the context of the DE model.

As usual, this criterion is expressed as a condition on the shape of
the stress signal of an underlying base state of homogeneous flow
response to the applied deformation. By definition, this base state
stress signal equals the globally measured one at least until any
significant banding takes place. Accordingly (\ref{eqn:
  step_strain_criterion}) can also be applied directly to the
experimentally measured stress signal.  (This assumes that no
appreciable banding developed during the ramp itself, which is a good
assumption for the fast finite-amplitude ramps of interest here: even
if the flow technically becomes linearly unstable to banding during
the ramp, there is insufficient time for heterogeneity to develop.)

\subsection{Numerical results: RP model}
\label{section: stress_relaxation}

\iffigures
\begin{figure}[tbp]
\includegraphics[width=8cm]{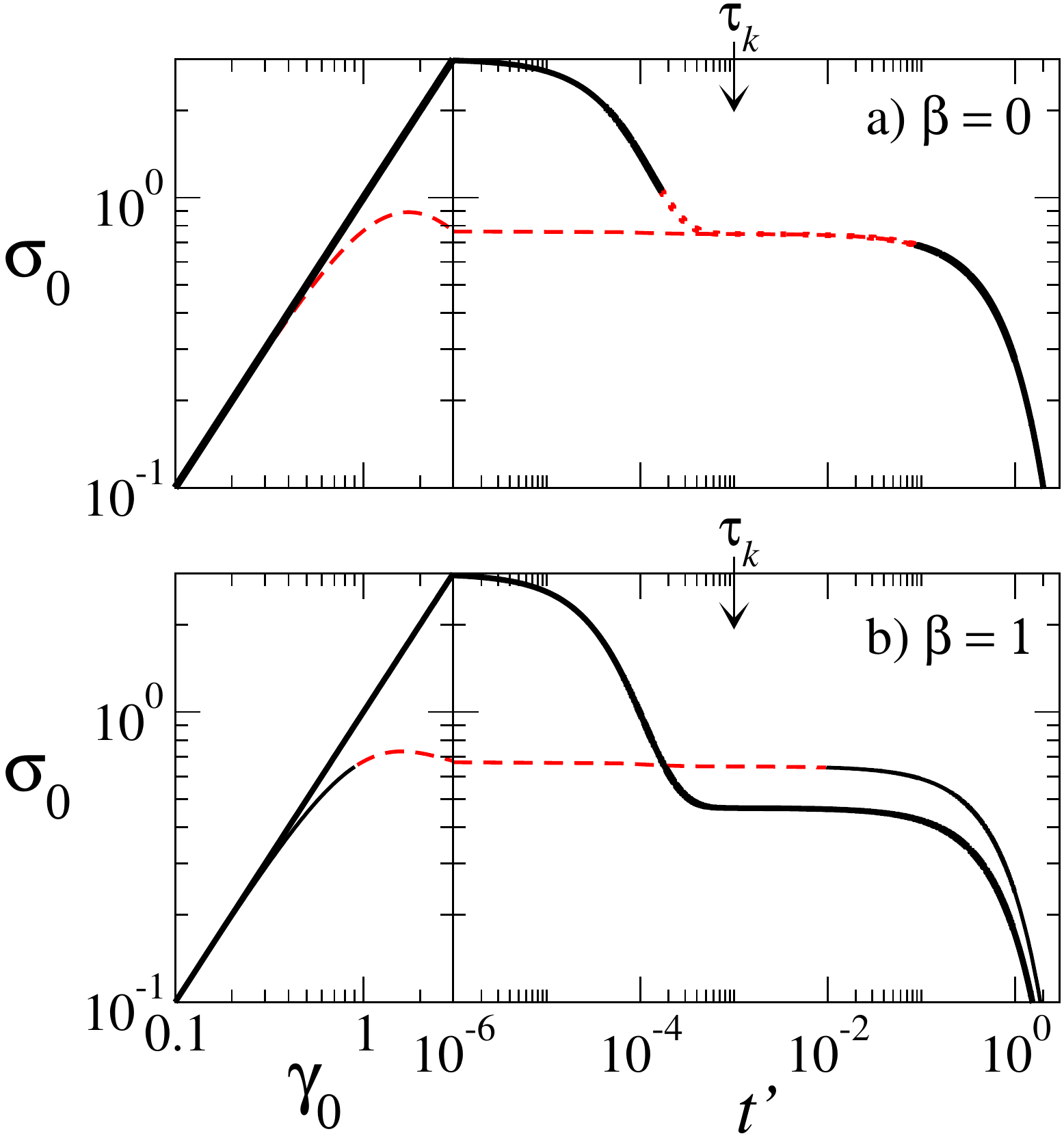}
\caption{Viscoelastic contribution to the shear stress in the RP model
  during (\vs $\gamma_0$) and after (\vs $t'$) a strain ramp of
  amplitude $\gamma^* = 3$, with homogeneity enforced. (a) $\beta = 0$
  (no CCR), (b) $\beta = 1$ (CCR active).  Dotted/dashed lines denote
  linearly unstable regions ($\omega > 0$).  Upper curve at the end of
  each ramp is for a ramp rate $\gdotsrp$ in the regime of chain
  stretch. Lower curve at the end of each ramp is for a ramp rate
  $\gdotnrp = 500$ in the non-stretch regime.  Parameters: $\taur =
  10^{-4}$, $\eta = 10^{-5}$, $\tauk = 10\taur$.}
 \label{fig: RP_relaxation}
\end{figure}
\fi

In the previous section above we developed a criterion for linear
instability, immediately following a rapid strain ramp, to the onset
of shear banding post-ramp. This criterion is expressed as a condition
on the shape of the stress signal of an underlying base state of
homogeneous flow response to the applied deformation, immediately as
the ramp ends.

In the next subsection below we shall present numerical results for
this base state signal throughout the full duration of its evolution,
both during and after ramp. We also present numerical results for any
regions of linear stability (negative eigenvalues) or instability
(positive eigenvalue) during this entire evolution. Recall that our
analytically derived criterion (\ref{eqn: step_strain_criterion})
above strictly only applies immediately post-ramp. In the second
subsection below we present the results of our spatially aware
nonlinear simulations of the shear bands that arise as the result of
any regimes of instability.

\subsubsection{Base state and linear instability}

Numerical results for the evolution of the base state stress signal,
both during and after the ramp, are shown for the RP model in
Fig.~\ref{fig: RP_relaxation}. (Here we have subtracted from
$\Sigma_0$ the trivial contribution $\eta\gdot_0$ from the Newtonian
solvent to leave the viscoelastic contribution $\sigma_0$ only.) For
clarity it is plotted as a function of strain during the ramp, and of
time afterwards. Regimes of linear instability to shear banding,
determined by numerically calculating the time-dependent eigenvalue of
the linearised equations discussed above, are shown as dotted and
dashed lines.  Consistent with our criterion (\ref{eqn:
  step_strain_criterion}) above, ramps that end with declining stress
leave the system unstable immediately post-ramp.

With this figure in mind, we discuss now in more detail separately
ramps that are slower and faster than the rate of stretch relaxation
$\invtaur$. (As noted above, in each case the ramp is faster than the
inverse terminal relaxation time, $\gdot_0\taud\gg 1$.)

Consider first a `slow' ramp at a rate $\gdotnrp$ for which $\gdotnrp
\taur \ll 1$.  For such a ramp, no appreciable chain stretch develops:
subscript `n' denotes nonstretching. The ramp is still nonetheless in
the fast flow regime $\gdot\taud\gg 1$ of the non-stretching version
of the model specified by Eqns.~\ref{eqn: nRP_components} above. The
corresponding mechanical response during the ramp can then by computed
by integrating only the terms prefactored by $\gdot$ in
Eqns.~\ref{eqn: nRP_components}: it is effectively that of a nonlinear
elastic solid with a stress signal that depends only on strain
$\base\Sigma=\base\Sigma(\base\gamma)$, independent of strain rate.
Numerical results for this are shown by the lower of the two curves
(during the ramp) in each panel of Fig.~\ref{fig: RP_relaxation}. It
displays an overshoot at an amplitude $\gamma_0 \sim 1.7$. The
system is therefore left unstable immediately post-ramp, as indicated
by the red dashed lines, consistent with our criterion (\ref{eqn:
  step_strain_criterion}).

Post-ramp the base state stress signal shows mono-exponential decay on
the single reptation timescale $\taud$ of tube reorientation, as the
system relaxes back to equilibrium.  Allied to this, the eigenvalue of
the stability analysis shows monotonic decay from its initial value
towards a final value $-( \frac{1}{\taud} + \frac{1}{\eta})$ which,
being negative, indicates stability of the final homogeneous
equilibrium state, as expected.  Any system left linearly stable
immediately post-ramp, by a ramp of amplitude $\gamma_0 < 1.7$, will
therefore remain stable for all subsequent times and exhibit no
banding. (This case is not shown in the figure.) In contrast, a system
that is left linearly unstable immediately post-ramp by a ramp of
amplitude $\gamma_0 > 1.7$, returns finally to a stable homogeneous
state. (See again the lower curve at the end of the ramp in each panel
of Fig.~\ref{fig: RP_relaxation}.) However shear bands do transiently
form during the relaxation process, as we shall discuss in more detail
in the next subsection below.

Having discussed `slow' ramps with $\gdotnrp \taur \ll 1$, we now
address `fast' ramps of typical rate $\gdotsrp$ such that $\gdotsrp
\taur \gg 1$. Here appreciable chain stretch develops during the ramp.
(Subscript `s' denotes stretching.) The associated mechanical response
during the ramp then follows by integrating the terms in $\gdot$ in the
sRP model Eqns.~\ref{eqn: sRP_components}. It again corresponds to
that of an elastic solid, independent of strain rate in this limit.
Indeed because of the simple linear structure of the $\gdot$ terms in
the sRP model, we now have a monotonically increasing relation
$\sigma_0=G\gamma_0$.  The system is therefore left linearly stable
against banding immediately post-ramp, as seen in the upper curve at
the end of the ramp in each panel of Fig.~\ref{fig: RP_relaxation},
and consistent with our analytical criterion (\ref{eqn:
  step_strain_criterion}).

\iffigures
\begin{figure*}[tbp]
  \centering
\includegraphics[trim = 2.5cm 6cm 2cm 6cm, clip = true, width=16cm]{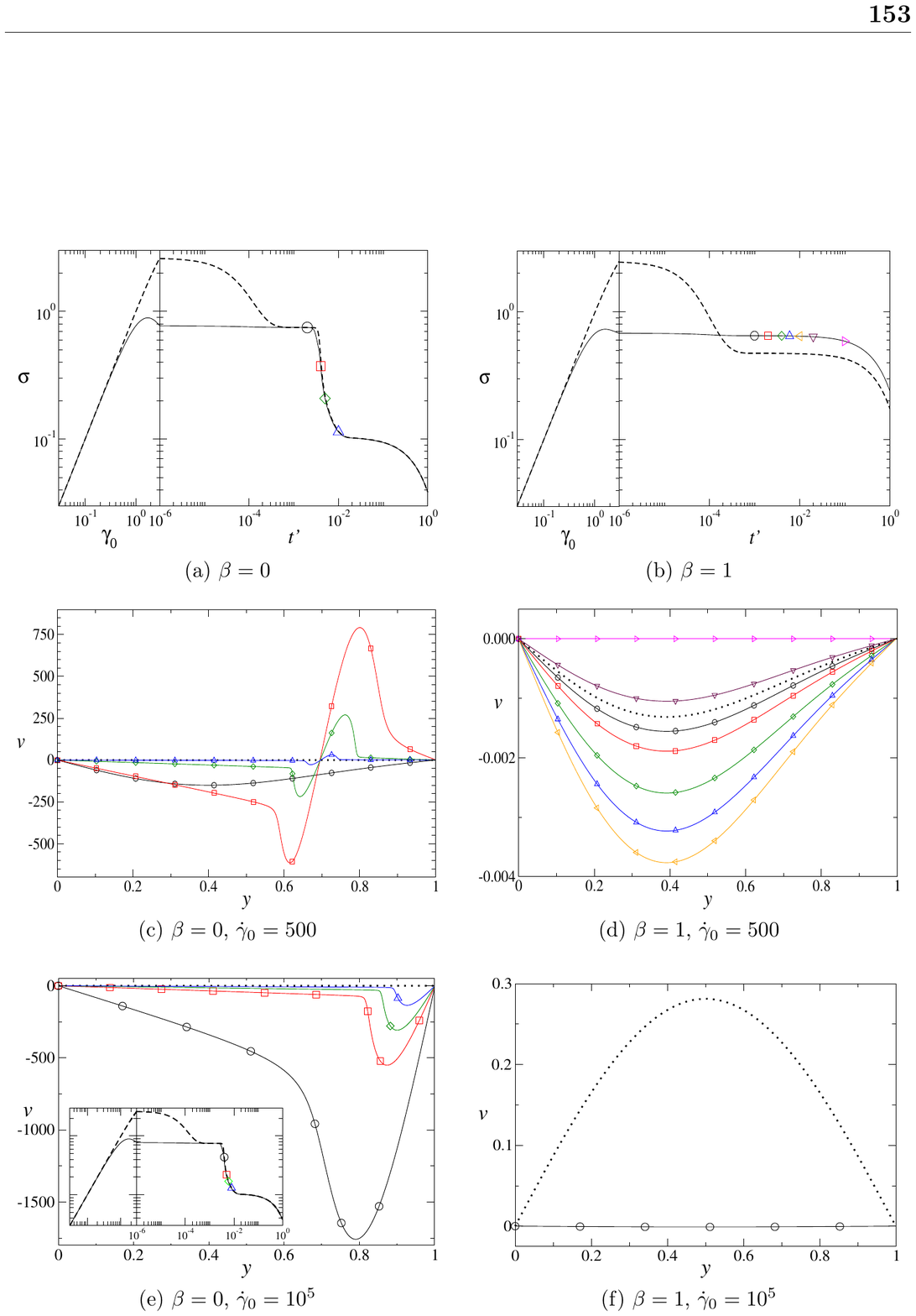}
\caption{Strain ramp in the RP model. (a) Shear stress during (\vs
  $\gamma_0$) and after (\vs $t'$) a ramp of amplitude $\gamma^* = 3$,
  with CCR inactive, $\beta = 0$.  The dashed line is for a ramp of
  fast rate $\gdotsrp = 10^5$ in the chain stretching regime; the
  solid line is for a ramp of rate $\gdotnrp = 500$ in the
  non-stretching regime.  (b) Corresponding curves with CCR active,
  $\beta=1$.  (c) and (d) show velocity profiles during stress
  relaxation at times corresponding to symbols in (a) and (b)
  respectively, in both cases following a ramp of rate in the
  non-stretching regime $\gdotnrp = 500$. Corresponding figures (e)
  and (f) are for a ramp of fast rate $\gdotsrp = 10^5$ in the
  stretching regime. Profiles are shown at times corresponding to
  symbols in the inset for (e), and at times corresponding to symbols
  in (b) for (f).  The normalised velocity heterogeneity $\base{v}(y)
  = v(y) - \gdot_0 y$ immediately before the end of the ramp at
  $t=t^{*-}$ is shown as a thick dotted line.  Model parameters $\taur
  = 10^{-4}$, $\eta = 10^{-5}$. Initial noise magnitude: $q = 5\times
  10^{-4}$.}
   \label{fig: hetero1}
\end{figure*}
\fi

However it is important to recall that our analytically derived
criterion (\ref{eqn: step_strain_criterion}) applies only
\emph{immediately} post-ramp and does not prescribe the system's
stability properties throughout the full duration of its relaxation
back to equilibrium post-ramp.  As seen in the upper curve (at the end
of each ramp) in each panel of Fig.~\ref{fig: RP_relaxation}, this
relaxation after a ramp of rate $\gdot_{0,s}\gg 1$ shows a double
exponential form: first as chain stretch relaxes on the fast timescale
$\taur$, and subsequently as tube reorientation takes place on the
much slower reptation timescale $\taud$.

Subfigures (a) and (b) respectively address a system without ($\beta =
0$) and with ($\beta = 1$) the mechanism convective constraint release
(CCR) active.  Immediately striking is the fact that, without CCR
(subfigure a) the first part of the stress relaxation on the fast
timescale of stretch relaxation $\taur$, returns the stress to a value
equal to that which would have been generated by a ramp of equivalent
amplitude in the slower non-stretching limit: the upper curve in
subfigure (a) rejoins the lower one on an intermediate plateau around
the time $\tau_k =10\taur$, before both finally follow the same decay
on the terminal timescale $\taud$.  Denoting by
$\sigma(t'=\tau_k,\gdot_0,\gamma_0)$ the stress on this intermediate
plateau, then, for a ramp of amplitude $\gamma_0$ we have
\beqn
\sigma(\tauk,\gdotsrp,\gamma_0) = \sigma(\tauk,\gdotnrp,\gamma_0) \quad \text{ for } \beta = 0.
\eeqn

Once this intermediate plateau has been attained, the stability
properties of the two curves in Fig.~\ref{fig: RP_relaxation}a)
coincide. Following a fast ramp of rate $\gdotsrp$ and amplitude
$\gamma_0 > 1.7$, therefore, we predict a \emph{delayed} banding
instability that sets in a time $O(\taur)$ post-ramp, even though no
stress overshoot occurred during the ramp itself.  This will be
confirmed by our spatially aware simulation showing shear banding in
the next subsection below. It is consistent with experimental results
that show delayed shear banding setting in on a timescale $O(\taur)$
post-ramp~\cite{Wangetal2009b, Archeretal1995a}.  


In contrast, with CCR active ($\beta \neq 0$) the stress remaining
after the initial part of the stress decay on the fast timescale
$\taur$ is significantly lower than that which would have been
generated by a ramp of corresponding amplitude in the slower,
non-stretching regime: the intermediate plateau values do not coincide
in Fig.~\ref{fig: RP_relaxation} (b).  Indeed, for large enough
$\beta$ this initial fast relaxation is sufficient to ensure that the
system remains stable against the formation of bands throughout the
full duration of its return to equilibrium, as seen in Fig.~\ref{fig:
  RP_relaxation}b). We therefore conclude that in order to observe
shear banding after a ramp in the chain stretching regime, the value
of the CCR parameter $\beta$ should be small $\beta \sim 0$.

These differences in the system's relaxation properties with and
without CCR can be explained as follows. Without CCR ($\beta = 0$),
the mechanisms of chain stretch relaxation and tube orientation
relaxation are decoupled and occur independently of each other. The
residual stress remaining after chain stretch has relaxed following a
`fast' ramp of rate $\gdotsrp$ is therefore equal to the stress that
would have resulted from a ramp of the same amplitude but rate
$\gdotnrp$ during which no chain stretch arose in the first place.  In
contrast, with CCR ($\beta \neq 0$) the relaxation of chain stretch
also brings significant relaxation in the orientation of tube
segments, because a proportion of the entanglements forming the tube
of constraints on a test chain are lost upon stretch relaxation. The
stress relaxation is thereby accelerated for times $t'<\taur$
compared with the non-CCR case.

\subsubsection{Nonlinear, spatially aware simulations}

So far, we have discussed the evolution of the base state stress
during and after a strain ramp, and its associated time-dependent
linear stability properties.  We now perform nonlinear simulations to
investigate the shear bands that form as a result of any regime of
linear instability. 

As can be seen in Fig.~\ref{fig: hetero1} the results are consistent
with our linear instability predictions of Fig.~\ref{fig:
  RP_relaxation}. Subfigures (c) and (d) show that shear rate
perturbations grow as soon as a ramp of rate $\gdotnrp$ and amplitude
$\gamma_0 > 1.7$ ends. For example, appreciable heterogeneity has
already developed by the time indicated by the circle in (a).  In
contrast, subfigures (e) and (f) show that after a `fast' ramp of rate
$\gdotsrp$ and amplitude $\gamma_0 > 1.7$, the system shows onset of
shear rate heterogeneity only after a delay time $t' \sim \tauk$, and
only for systems in which CCR is sufficiently small $\beta \sim 0$
(subfigure e).  With CCR active (subfigure f) any residual
heterogeneity at the end of the ramp decays monotonically.

Fig.~\ref{fig: hetero1} (a) also demonstrates that shear rate
heterogeneity of the large magnitude seen in this protocol can
dramatically alter the stress relaxation function. As the local shear
rate becomes extremely large, nonlinearities become important and
result in a sudden and dramatic acceleration of stress relaxation
compared with the base state signal of Fig.~\ref{fig: RP_relaxation}.
This causes the second drop-off in stress in that subfigure. (Recall
that the first drop-off after the fast ramp in contrast arose from
chain stretch relaxation in the underlying base state.)

\subsection{Comparison with experiment}

\iffigures
\begin{figure}[tbp]
  \centering
\includegraphics[width=7cm,height=5cm]{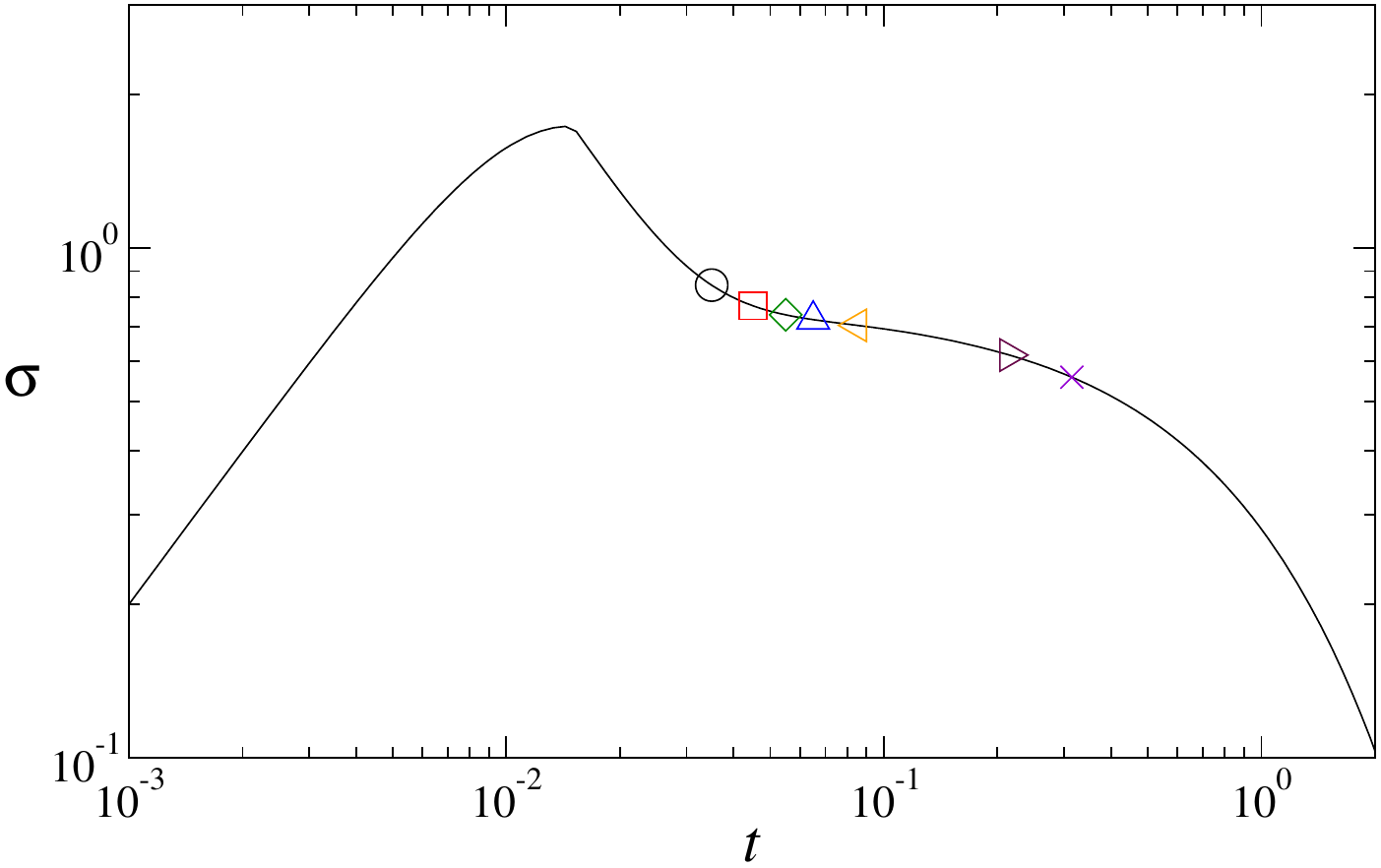}
   \caption{Viscoelastic stress during and after a strain ramp of amplitude $\gamma_0 =3$ and rate $\gdot_0 = 200$, time of shear cessation: $t^* = 0.015$. Here $\taur = 10^{-2}$, $\eta = 10^{-5}$, $\beta = 0$, and initial noise magnitude: $q= 10^{-2}$.}
\label{fig: hetero2a}
\end{figure}

\begin{figure}[tbp]
  \centering
\includegraphics[width=7cm,height=5cm]{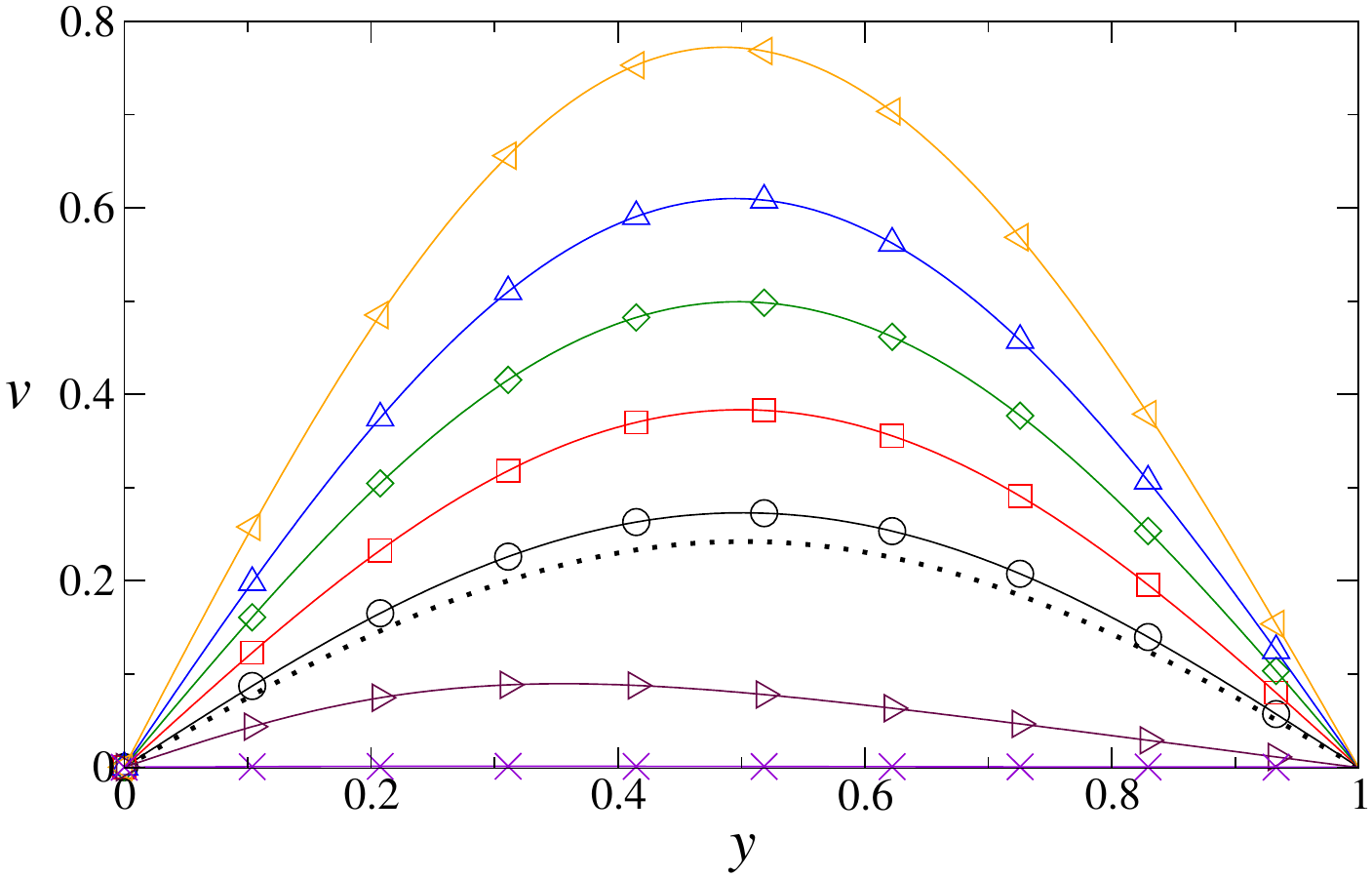}
\caption{Snapshots of the velocity profile during stress relaxation
  after a ramp shown in Figure \ref{fig: hetero2a}, at times
  corresponding to symbols in that figure. Dotted line: snapshot of
  the normalised velocity profile $\base{v}(y)= v(y) - \gdot_0 y$ at
  time $t = t^{*-}$.}
\label{fig: hetero2b}
\end{figure}
\fi

We now discuss our results in relation to experimental observations of
shear banding following the imposition of a fast strain ramp in
polymeric fluids. In doing so, we recall that shear banding is often
described as `macroscopic motions' in the context of this protocol.

For strain ramps that terminate in a regime of declining stress versus
strain, macroscopic motions accompanied by a dramatic drop in the
stress signal were reported to develop quickly post-ramp in
Refs.~\cite{Wangetal2008c, Wangetal2009c}.  This is consistent with
our analytical criterion (\ref{eqn: step_strain_criterion}), and also
with our numerics in Figs.~\ref{fig: hetero1}c) and d).

For strain ramps of amplitudes $\gamma_0 \gtrsim 1.5$ that show no
stress overshoot during the ramp, delayed macroscopic motions have
been reported post-ramp in polymer melts~\cite{Wangetal2009b,
  Wangetal2009c} and solutions~\cite{Archeretal1995a}. These onset
after a time $t'=O(\taur)$ and are concurrent with a sudden drop
in the stress signal. These observations are consistent with the
stress response and associated transient shear banding in the RP model
for $\beta \sim 0$ following a strain  ramp of rate $\gdotsrp$ and
amplitude $\gamma_0 \gtrsim 1.7$: recall Fig.~\ref{fig: hetero1}e).

Experimental reports of similar macroscopic motions but without the
accompanying dramatic drop in the stress signal post-ramp are also
widespread~\cite{Wangetal2006a, Fangetal2011a, Wangetal2009b,
  Wangetal2010a, Wangetal2010b, Wangetal2007a}. Qualitatively similar
behaviour can be uncovered in the RP model by decreasing the
entanglement number $Z$ to give less well separated relaxation
timescales $\taud, \taur$. This decreases the maximal degree of
banding observed post-ramp to a sufficient extent that the global
stress signal no longer differs significantly from that of the
homogeneous constrained system. An example of such behaviour in shown
in Figs.~\ref{fig: hetero2a} and~\ref{fig: hetero2b}. The stress
signal now shows a single dropoff associated with chain stretch
relaxation, but lacks a second dropoff associated with the nonlinear
effects of banding.

Indeed, our numerical runs in Figs.~\ref{fig: RP_relaxation}
and~\ref{fig: hetero1} assumed an entanglement number $Z=3300$ larger
than is the case experimentally. We set this number deliberately to
ensure a clean separation of timescales $\taud=10^4\taur$, thereby
allowing a clear pedagogical discussion of the relative effects of the
relaxation of chain stretch and the relaxation of tube orientation
post-ramp.

In practice, however, polymer melts with $Z \gtrsim 100$, and DNA or
wormlike micelles with $Z \gtrsim 150$, usually suffer edge fracture
so that reliable results are difficult to obtain.  The separation of
time scales in experiment is therefore more typically $\taud/\taur
\sim 150 - 500$.  Upon repeating our simulations for these more
realistic values of $\taud/\taur$, and for less severe ramp rates
$\gdotsrp$, we find, reassuringly, that the qualitative features
described above still obtain (provided $\taud/\taur \gtrsim 10$ and
$\gdotsrp > \invtaur$). See Figs.~\ref{fig: hetero2a} and~\ref{fig:
  hetero2b}. We do not attempt to reproduce specific experimental
results by matching the relaxation times or ramp rates in detail in
this work, because studies elsewhere have done
so~\cite{Agimelenetal2012a,AgimelenThesis}.

Finally, we comment on our numerical findings in relation to the A, B,
C classification system discussed in Sec.~\ref{sec:observations}
above. As seen towards the right hand side of Figs.~\ref{fig:
  hetero1}, the development of very large local shear rates post-ramp
is closely associated with a stress relaxation that is accelerated
compared with that which would be predicted by a calculation in which
homogeneity is imposed by assumption. This can result in a significant
decrease in the damping function compared with that predicted by any
homogeneous calculation and so to type C behaviour.  Conversely, for
systems in which the macroscopic motions that develop are more modest
(even if still observable experimentally), the stress relaxation
function agrees well with that of a calculation in which homogeneity
is assumed, leading to a damping function of type A.

Furthermore, a decreasing separation of relaxation times $\taud/\taur$
(decreasing entanglement number $Z$) causes the maximal degree of
banding to decrease and so could lead correspondingly to a progression
from type C to type A behaviour. This is consistent with reports that
type C behaviour is most common in very well entangled polymers, while
type A behaviour occurs most often for moderately entangled polymers
\cite{Venerus2005a, Osaki1993a}.  We conclude that the RP model is
capable of exhibiting both types A and C behaviour, with a progression
between the two consistent with that seen experimentally.  To the best
of our knowledge, the RP model is unable to show the less commonly
reported type B behaviour of very weakly entangled
materials~\cite{Venerus2005a, Osaki1993a}. Indeed, the RP model is in
any case not aimed at describing these systems.

We note finally that the Giesekus model has a linear stress-strain
relation in a fast strain ramp protocol: $\Sigma \sim G\gamma$. It
therefore predicts linearly stability against shear banding
immediately after a fast strain ramp, according to (\ref{eqn:
  step_strain_criterion}). Furthermore this model contains only one
relaxation time. Accordingly it is unlikely to capture the rich
experimental phenomenology just discussed for this protocol, and we do
not discuss it further here.


\section{Results: shear startup}
\label{section: shearstartup}

We consider finally the shear startup protocol, in which a previously
well rested sample is subject to shearing at a constant rate $\gdot_0$
for all times $t>0$, giving a strain $\gamma_0=\gdot_0 t$.  We first
derive an analytical criterion for the onset of banding in this
protocol, before presenting numerical results that support it.

\subsection{Criterion for instability in shear startup}

In a shear startup experiment, the most commonly reported rheological
response function is that of the startup stress as a function of time
$t$, or equivalently as a function of strain $\gamma_0=\gdot_0 t$, for
the given applied strain rate $\gdot_0$. When plotted as a function of
accumulated strain $\gamma_0$ for a collection of startup runs, each
performed at a constant value of the strain rate $\gdot_0$, this gives
us a two-dimensional function $\Sigma_0(\gamma_0,\gdot_0)$.

In the context of shear banding, a familiar thought-experiment is to
consider an (artificial) situation in which a startup flow is
constrained to remain homogeneous until the system attains a
stationary state in the limit $\gamma_0\to \infty$. In this limit, the
total accumulated strain becomes irrelevant and the stress depends
only on strain rate: $\Sigma_0=\Sigma_0(\gdot_0)$. The criterion for
shear banding (with the constraint now removed) is well known in this
limit: that the underlying constitutive curve of stress as a function
of strain rate has negative slope, $d\Sigma_0/d\gdot_0 < 0$.

Our aim here is to generalise this result, which is valid only for a
stationary homogeneous base state in the limit $\gamma_0\to\infty$, to
finite strains $\gamma_0$ and times $t$, in order to predict at what
stage during startup banding first sets in. As we shall show, the
onset of banding is closely associated with the time (or equivalent
strain) of any overshoot in the stress startup signal
$\Sigma_0(\gamma_0)$, for a given applied strain rate $\gdot_0$.
Clearly this implies an onset criterion $d\Sigma_0/d\gamma_0<0$, which
is indeed a very useful rule of thumb to apply to experimental data.
However we show below that it is in fact modified slightly, leading to
onset a little {\rm before} overshoot.

Besides predicting the time at which bands first start to form in any
experiment for which the eventual steady state is banded,
$d\Sigma_0/d\gdot_0<0$, an important further outcome of what follows
will be to predict the transient appearance of shear bands, again
associated with a startup overshoot and subsequently declining stress
$d\Sigma_0/d\gamma_0<0$, even in systems for which the underlying
constitutive curve is monotonic $d\Sigma_0/d\gdot_0 > 0$ and the
steady state unbanded.

As ever, our strategy will be to consider an underlying base state of
initially homogeneous flow response, and the dynamics of small
heterogeneous perturbations about it. Our criterion for the growth of
these perturbations, {\it i.e.}, for the onset of banding, will be
expressed in terms of the partial derivatives of the base state stress
signal $\Sigma_0(\gamma_0,\gdot_0)$ with respect to $\gamma_0$ and
$\gdot_0$.

In many places below we shall graphically present results in the plane
of $\gamma_0$ and $\gdot_0$.  To interpret data presented in this way,
it is useful to keep in mind that a vertical cut up this plane at its
far right hand side $\gamma_0\to\infty$ corresponds to the system's
steady state properties as a function of strain rate $\gdot_0$.  A
horizontal cut corresponds to the system's startup behaviour as a
function of accumulated strain $\gamma_0$, in a single startup run
performed at a fixed value of the strain rate $\gdot_0$.

As noted in the context of the other protocols above, because the base
state signal corresponds to the experimentally measured one at least
until appreciable banding develops, the criterion that we develop can
be applied directly to experimentally measured stress startup data.

To start, then, we consider the properties of an underlying base state
of initially homogeneous flow response to an imposed shear startup
deformation. Were the flow to remain homogeneous through to the
stationary limit $\gamma_0\to\infty$, the condition for banding
instability would then be that the stationary constitutive curve of
stress as a function of strain rate has a region of negative slope
$d\Sigma_0/d\gdot_0 < 0$, as noted above. In practice, however, the
flow generally becomes unstable to banding before this stationary
limit is attained.  As a first step to generalising our onset
criterion to finite strains during startup, we define a fixed-strain
constitutive curve:
\beqn 
\Sigma_0(\gdot_0)|_{\gamma_0={\rm const.}}=G W_{xy0}(\gdot_0)|_{\gamma_0={\rm const.}}+\eta\gdot_0.
\eeqn
Experimentally, such a curve would be constructed by performing a
series of startup runs at different shear rates and plotting the shear
stress, grabbed at the same fixed strain $\gamma_0$ in each run, as a
function of the applied shear rate.

We then consider the derivative of this fixed-strain constitutive
curve with respect to shear rate:
\beqn
\label{eqn: diff_fixed}
\partial_{\gdot_0}\base{\Sigma}|_{\base{\gamma}} = G\partial_{\gdot_0}\base{{W}_{xy}}|_{\gamma_0}+\eta.
\eeqn
This clearly reduces to the slope of the underlying steady state
constitutive curve $d\Sigma_0/d\gdot_0$ in the limit $\gamma_0\to
\infty$, and more generally is the finite-strain analogue of it.

To proceed further, we need an expression for
$\partial_{\gdot_0}\base{{W}_{xy}}|_{\gamma_0}$.  To obtain this we
return to Eqn.~\ref{eqn: governing_eqn_diffusive}, divided across by
strain rate:
\beqn
\partial_{\gamma_0\,} \vecv{\base{s}}|_{{\base{\gdot}}} =  \frac{1}{\base{\gdot}}\vecv{Q}(\vecv{\base{s}},\base{\gdot}).
\label{eqn: crit1}
\eeqn
(We again neglect diffusive terms, which are small for the most
unstable mode in the linear regime.) Differentiating this with respect
to strain rate gives
\beqn
\partial_{{\gdot_0}\,}\partial_{{\gamma_0}\,}\vecv{{\base{s}}}= -\frac{1}{\base{\gdot}}\partial_{{\gamma_0}\,}\vecv{{\base{s}}}|_{{\base{\gdot}}}+\frac{1}{\base{\gdot}}\tens{M}\cdot \partial_{{\gdot_0}\,}\vecv{{\base{s}}}|_{\gamma_0}+\frac{1}{\base{\gdot}}\vecv{q},
\eeqn
in which $\tens{M} =
\partial_{\vecv{s}\,}\vecv{Q}\mid_{\vecv{{\base{s}}},{\base{\gdot}}}$
and $\vecv{q} =
\partial_{\gdot}\vecv{Q}\,|_{\vecv{\base{s}},{\base{\gdot}}}$, as
previously. Multiplying by $\gdot_0\tens{M}^{-1}$ and rearranging we have
\beqn
\partial_{{\gdot_0}\,}\vecv{{\base{s}}}|_{\gamma_0}=\tens{M}^{-1}\cdot\left(\partial_{{\gamma_0}\,}\vecv{{\base{s}}}|_{{\base{\gdot}}}-\vecv{q}+\gdot_0\partial_{{\gdot_0}\,}\partial_{{\gamma_0}\,}\vecv{{\base{s}}}\right).
\eeqn
Using $\vecv{p}$ to project out the first component gives
\beqn
\partial_{\gdot_0}\base{{W}_{xy}}|_{\gamma_0}=\vecv{p}\cdot\tens{M}^{-1}\cdot\left(\partial_{{\gamma_0}\,}\vecv{{\base{s}}}|_{{\base{\gdot}}}-\vecv{q}+\gdot_0\partial_{{\gdot_0}\,}\partial_{{\gamma_0}\,}\vecv{{\base{s}}}\right),
\eeqn
which, substituted into Eqn.~\ref{eqn: diff_fixed}, gives finally an
expression
\beqn
\partial_{{\gdot_0}}{\base{\Sigma}}|_{{\base{\gamma}}} = G\,\vecv{p} \cdot \tens{M}^{-1} \left( \partial_{{\gamma_0}\,}\vecv{{\base{s}}}|_{{\base{\gdot}}}-\vecv{q}+\gdot_0\partial_{{\gdot_0}\,}\partial_{{\gamma_0}\,}\vecv{{\base{s}}} \right) + \eta
\label{eqn: crit2}
\eeqn
for the derivative with respect to shear rate of the fixed-strain
constitutive curve of the underlying homogeneous base state. We shall
return to this expression in a few lines below.

We now turn to consider the dynamics of any heterogeneous
perturbations to the homogeneous base state just discussed.  Recalling
Eqns.~\ref{eqn: one} and~\ref{eqn: two}, we have
\beqn
\partial_{t\,}\vecv{\delta s_n} = \left(\tens{M} -\frac{G}{\eta}\vecv{q}\,\vecv{p}\right) \cdot \vecv{\delta s_n}.
\label{eqn: crit3}
\eeqn
The criterion for this system of linear equations to have a
positive eigenvalue, which signifies onset of instability to the
growth of shear banding perturbations at any time, is
\beqn
(-1)^D\left|\tens{M} - \frac{G}{\eta} \vecv{q}\,\vecv{p}\right| > 0,
\eeqn
where $D$ is the dimensionality of $\tens{M}$. (We neglect the
possibility of the emergence of two complex conjugate eigenvalues of
positive real part - a Hopf bifurcation - because we have never seen
this in practice in our numerics.) This corresponds exactly to
\beqn
(-1)^D\left|\tens{M}\right|(
1-\frac{G}{\eta}\vecv{p} \cdot \tens{M}^{-1} \cdot \vecv{q})>0.
\eeqn
Using the fact that $(-1)^D\left|\tens{M}\right|<0$ (which follows
from noting that the base state must be stable with respect to {\em
  homogeneous} perturbations at fixed $\gdot_0$), this further
corresponds exactly to
\beqn
\eta-G\vecv{p} \cdot \tens{M}^{-1} \cdot \vecv{q}<0.
\eeqn
Combining this with equation (\ref{eqn: crit2}) above for the base
state, we find finally an exact criterion for the onset of a linear
instability to shear banding during startup:
\beqn
\partial_{\gdot_0}\base{\Sigma}|_{\base{\gamma}} - G\,\vecv{p} \cdot \tens{M}^{-1} \cdot (\partial_{\gamma_0}\vecv{\base{s}}|_{\base{\gdot}} + \base{\gdot} \partial_{\gdot_0\,}\partial_{\gamma_0\,}\vecv{\base{s}}) < 0.
\label{eqn: crit_default}
\eeqn

In raw form, this criterion appears cumbersome and somewhat removed
from conveniently measurable experimental quantities.  However its
overall structure is physically transparent. The first term is a
derivative of the base state stress with respect to strain rate.  The
second term is a derivative of the base state with respect to strain.
The third is a cross term, containing derivatives with respect to
both. To illuminate its physical content, therefore, we start by
discussing two distinct and physically important limits in which the
first and second terms separately dominate.

Consider first an (artificial) situation in which a homogeneous
startup flow proceeds through to a stationary state in the limit of
large strain $\gamma_0\to\infty$, without banding en route. In this
limit, derivatives with respect to strain $\partial_{\gamma}$ vanish
from (\ref{eqn: crit_default}) and we recover the familiar, and much
simpler, criterion for banding in steady state already discussed above
\beqn
\label{eqn: crit_viscous}
\partial_{\gdot}\base{\Sigma}|_{\base{\gamma} \to \infty}<0.
\eeqn
This criterion also applies (less artificially) to the onset of a
linear instability to shear banding during an experiment in which the
strain rate $\gdot_0$ is slowly swept upwards from zero. Because the
material is flowing in a liquid-like way in this limit, we term this a
`viscous instability' for convenient nomenclature in what follows.

Consider conversely a single startup run performed in the limit of a
very fast flow $\gdot_0 \to \infty$. In this regime many viscoelastic
materials behave essentially as elastic solids, with the stress
startup function converging to a limiting curve
$\base{\Sigma}(\base{\gamma})$ from which any dependence on shear rate
is lost, $\partial_{\gdot}\to 0$.  The full criterion~(\ref{eqn:
  crit_default}) then reduces to
\beqn
-G\,\vecv{p} \cdot \tens{M}^{-1} \cdot \partial_{\gamma_0}\vecv{s}\,|_{\gdot_0} < 0.
\label{eqn: crit_elas}
\eeqn
Because in this limit the material responds essentially as an elastic
solid, we term this an `elastic instability' for convenient
nomenclature in what follows.

Although simpler than (\ref{eqn: crit_default}), (\ref{eqn:
  crit_elas}) is still not expressed in terms of quantities that are
easily measured experimentally.  However further simplification is
possible in the case of only two dynamical variables $D=2$, for
example in flow regimes in which the dynamics is dominated by the
shear stress and only one component of normal stress difference.  In
this case (\ref{eqn: crit_elas}) further reduces to
\beqn
\label{eqn: crit_elas_2dof}
-\frac{1}{\gdot_0^2}\text{tr}\tens{M}\, \partial_{\gamma_0}\base{\Sigma}|_{\base{\gdot}} +  \frac{1}{\gdot_0}\partial_{\gamma_0}^{2}\,\base{\Sigma}|_{\base{\gdot}} < 0\;\;\;\rm{with}\;\;\;\text{tr}\tens{M}<0.
\eeqn

Taken alone, the first term of this expression predicts onset of
banding immediately after any overshoot in the stress as a function of
strain during startup. The second term modulates this result slightly,
allowing onset slightly before overshoot, once the stress starts to
curve downwards. This prediction is consistent with numerous
experimental observations of time-dependent shear banding associated
with stress overshoot during startup: in soft glassy
materials~\cite{Divouxetal2010a, Divouxetal2011a} and entangled
polymer melts and solutions~\cite{ Wangetal2008a, Wangetal2008b,
  Wangetal2009a, Wangetal2006b, Huetal2007a}, and also in simulation
studies~\cite{Zhouetal2008a, Adamsetal2009a, Adamsetal2009b,
  Adamsetal2011, Moorcroftetal2011, Likhtmanetal2012,
  Manningetal2007a}.

\subsection{Numerical results: rolie-poly model}
\label{section: nRP_startup}

In this section, we present our numerical results for shear startup in
the RP model. We consider first the limit in which polymer chain
stretch is negligible, $\gdot_0\taur\ll 1$, before commenting on the
effects of stretch.

The behaviour of the rolie-poly model in startup has been studied
previously numerically in Refs.~\cite{Adamsetal2011,Adamsetal2008a}.
One of our aims in what follows is to understand the phenomena
reported in that work, some of which we must necessarily reproduce in
our numerics here, in the context of the general analytical criterion
developed above.

\subsubsection{Nonstretching rolie-poly model}

\iffigures
\begin{figure}[tbp]
  \includegraphics[width=8cm,height=5cm]{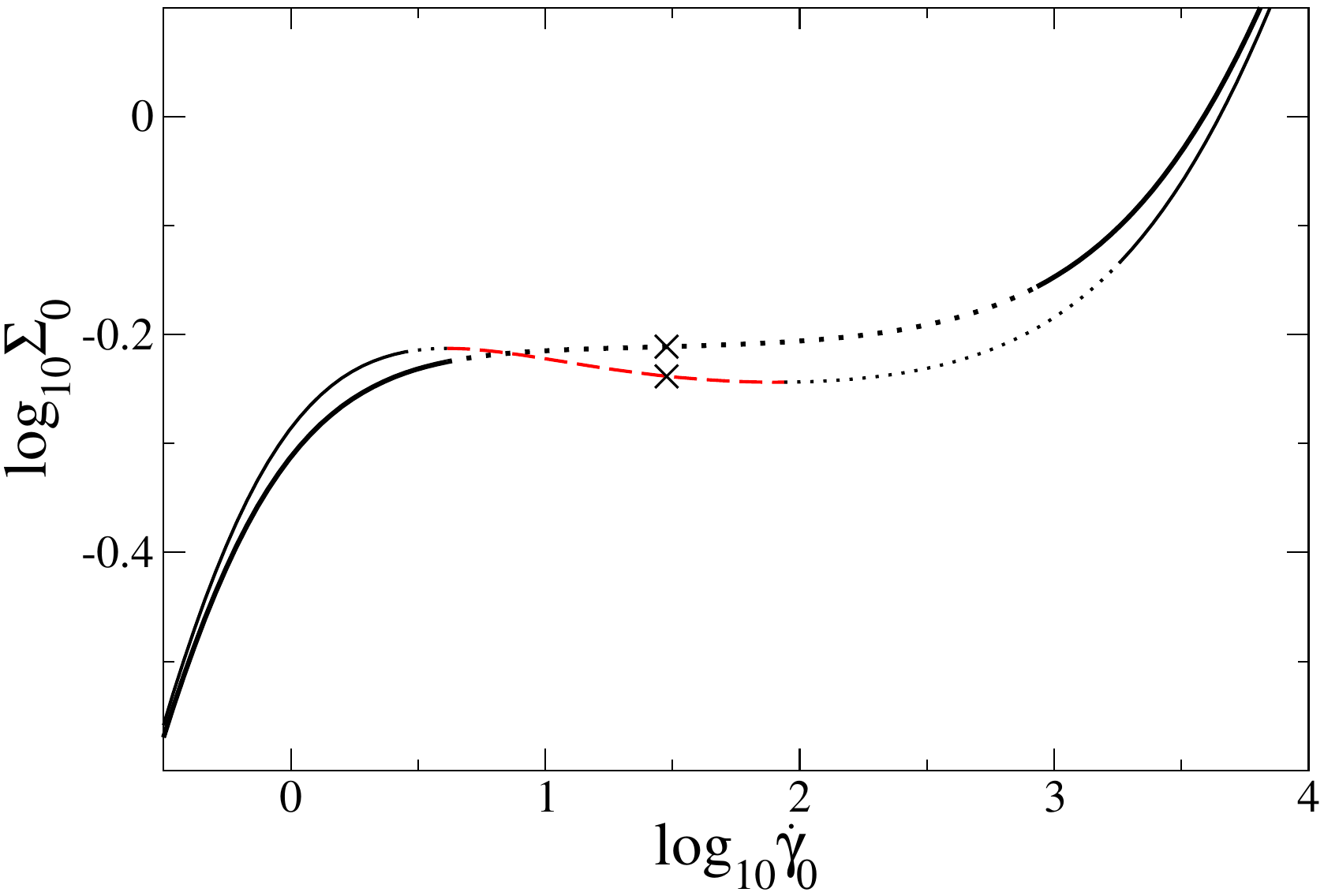}
  \caption{Constitutive curves of the nRP model for $\beta = 0.4,1$
    (bottom to top on the right) and $\eta = 10^{-4}$. Dashed:
    linearly unstable at steady state. Dotted: \emph{transiently}
    linearly unstable before the steady state is reached. Crosses
    denote shear rate $\gdot = 30$ for which time-dependent shear
    startup behaviour is explored in Fig.~\ref{fig:
      RP_example_transient}.}
  \label{fig: RP_constitutive_curves}
\end{figure}
\fi

\iffigures
\begin{figure*}[tbp]
\includegraphics[width=16cm]{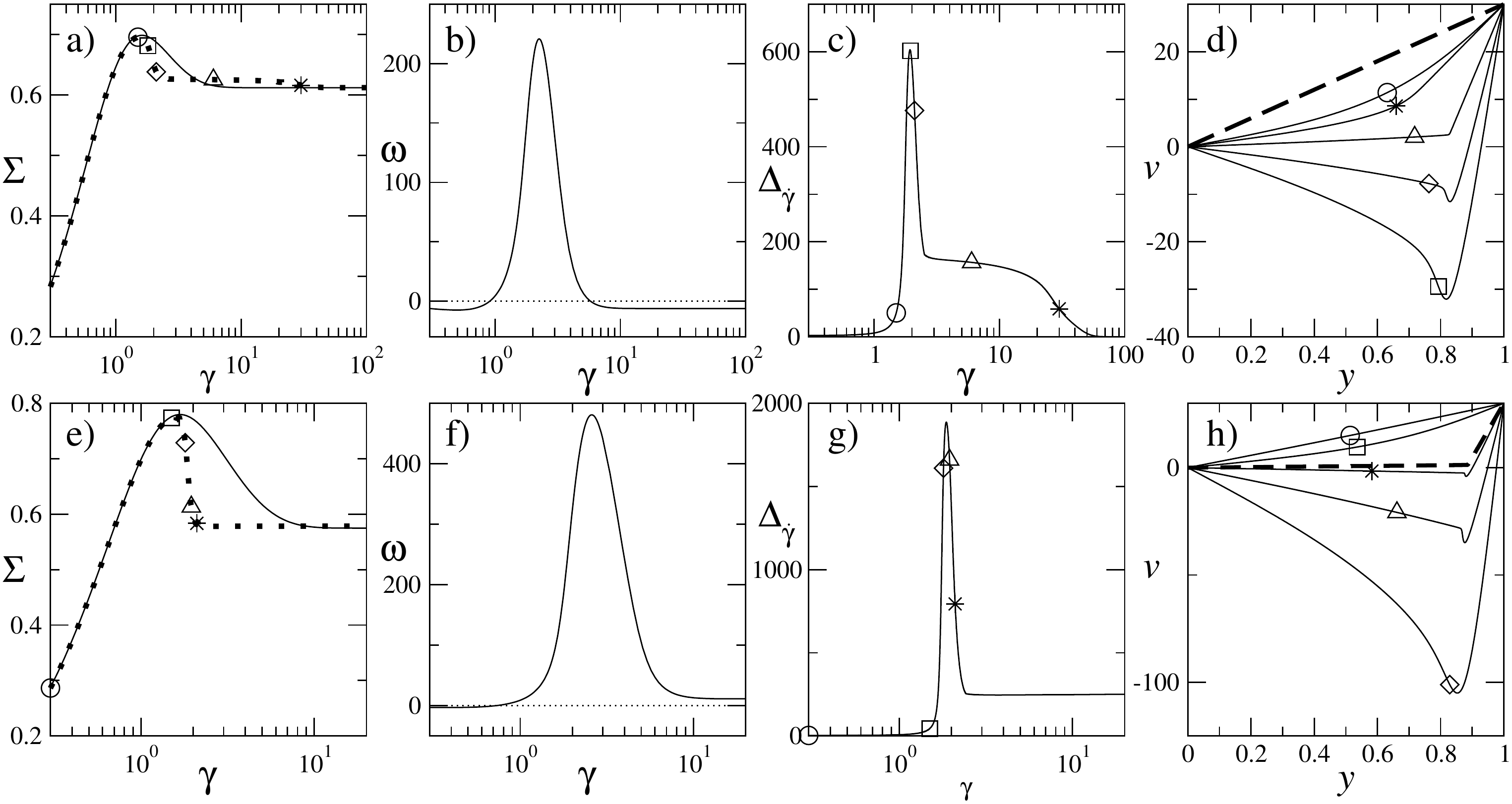}
\caption{Responses of the nRP model to an imposed shear rate $\gdotbar
  = 30$ for the case of a monotonic constitutive curve $\beta = 1$
  (top row) and for the case of non-monotonic constitutive curve
  $\beta=0.4$ (bottom row). (a, e) Shear stress response with
  homogeneity enforced (solid line) and with heterogeneity allowed
  (dotted line). (b, f) Linear stability analysis results for the real
  part of the eigenvalue that has the largest real part. (c, g).
  Degree of banding $\Delta_{\gdot} = \gdot_{\rm max}-\gdot_{\rm min}$
  in full nonlinear simulations . (d, h) Snapshots of the velocity
  profile at strains corresponding to symbols in a/c), e/g). Steady
  state velocity profile is shown as a thick dashed line.  $\eta =
  10^{-4}$, $q=0.1$.}
  \label{fig: RP_example_transient}
\end{figure*}
\fi

Depending on the values of the model parameters $\beta,\eta$, the
underlying constitutive curve of the non-stretch rolie-poly (nRP)
model can either be monotonic or non-monotonic.  A representative
example of each case is shown in
Fig.~\ref{fig: RP_constitutive_curves}.

We consider first the non-monotonic case, for a shear rate indicated
by the cross in the negatively sloping regime $d\Sigma_0/d\gdot_0<0$.
The model's shear startup behaviour at this imposed shear rate is
explored in the bottom row of Fig.~\ref{fig: RP_example_transient}.
The velocity profile shown by the thick dashed line in the bottom
right subfigure shows that the steady flowing state is shear banded,
consistent with a `viscous instability' implied by the negative slope
$d\Sigma_0/d\gdot_0<0$ in the constitutive curve.

Also immediately obvious in Fig.~\ref{fig: RP_example_transient} is
the fact that bands first form rather early during the startup
process, apparently triggered by an `elastic instability' associated
with the startup stress overshoot and subsequently declining stress
$d\Sigma_0/d\gamma_0<0$.  When first formed these are much more
pronounced than in steady state, as shown by the pronounced spike in
the degree of banding $\Delta_{\gdot}$. Indeed this `elastic banding'
can be so pronounced as to precipitate negative local shear rates and
negative local velocities at some regions across the cell, consistent
with the material behaving as an elastic solid subject to a declining
stress in this regime. Clearly, then, in shear startup an `elastic
instability' associated with stress overshoot can precede and be much
more violent than any `viscous instability' associated with steady
state banding.

For shear rates shown by the dotted lines either side of the
negatively sloping regime $d\Sigma_0/d\gdot_0<0$ in
Fig.~\ref{fig: RP_constitutive_curves}, steady state `viscous
instability' is absent, but a pronounced `elastic instability' can
nonetheless still arise during startup. This leads to the formation of
pronounced banding that persists only transiently, before decaying to
leave homogeneous flow in steady state.

The model's startup behaviour across a wide range of imposed shear
rates $\gdot_0$ is summarised in Fig.~\ref{fig:
  portrait_noiseinitonly_beta1}. In the left panel of this figure we
show our linear stability criteria for the onset of banding in the
full plane of $\gdot_0,\gamma_0$.  As noted above, horizontal cut
across this plane corresponds to the fluid's startup behaviour as a
function of accumulated strain, at a single fixed value of the strain
rate.  A vertical cut at the far right hand side corresponds to the
system's steady state properties as a function of strain rate.  We
again use parameter values for $\beta,\eta$ corresponding to the
non-monotonic constitutive curve in Fig.~\ref{fig:
  RP_constitutive_curves}.  Accordingly, the startup run explored in
the bottom row of Fig.~\ref{fig: RP_example_transient} corresponds to
horizontal slice through Fig.~\ref{fig: portrait_noiseinitonly_beta0.4}
at a fixed value of $\gdot_0 = 30$.

In the left panel of Fig.~\ref{fig: portrait_noiseinitonly_beta0.4},
then, the black dotted line indicates the locus of strain values
$\gamma_0$ for which the base state stress startup curves show a
stress overshoot, with these curves measured across a range of closely
spaced values of $\gdot_0$. In other words, in any single startup run
corresponding to a horizontal cut across this plane at a fixed
$\gdot_0$, the stress overshoot occurs at the strain indicated by this
black dotted line.  This line being vertical indicates that in fact
the stress overshoot occurs at a fixed strain $\gamma \sim 1.7$ for
all values of imposed shear rates.

The green solid line indicates the strain at which our criterion
(\ref{eqn: crit_elas_2dof}) for `elastic instability' is first met in
each horizontal startup slice.  As can be seen this occurs just
before overshoot in each run, due to the presence of the stress
curvature terms in (\ref{eqn: crit_elas_2dof}).

The dashed line encloses the region of viscous instability in which
$\partial_{\gdot}\base{\Sigma}\mid_{\base{\gamma}} < 0$, according to
our criterion~(\ref{eqn: crit_viscous}). At the far right hand side of
the plane $\gamma_0\to\infty$ this coincides with the region of
negative slope in the underlying constitutive curve.

The large open circles enclose the region in which the full
criterion~(\ref{eqn: crit_default}) for linear instability to banding
is met.  We have cross-checked numerically that this indeed coincides
with the region in which there exists a positive eigenvalue of the
linearised equations, thereby verifying our analytical derivation
of~(\ref{eqn: crit_default}).

\iffigures
\begin{figure}[tbp]
 \includegraphics[width=8.5cm]{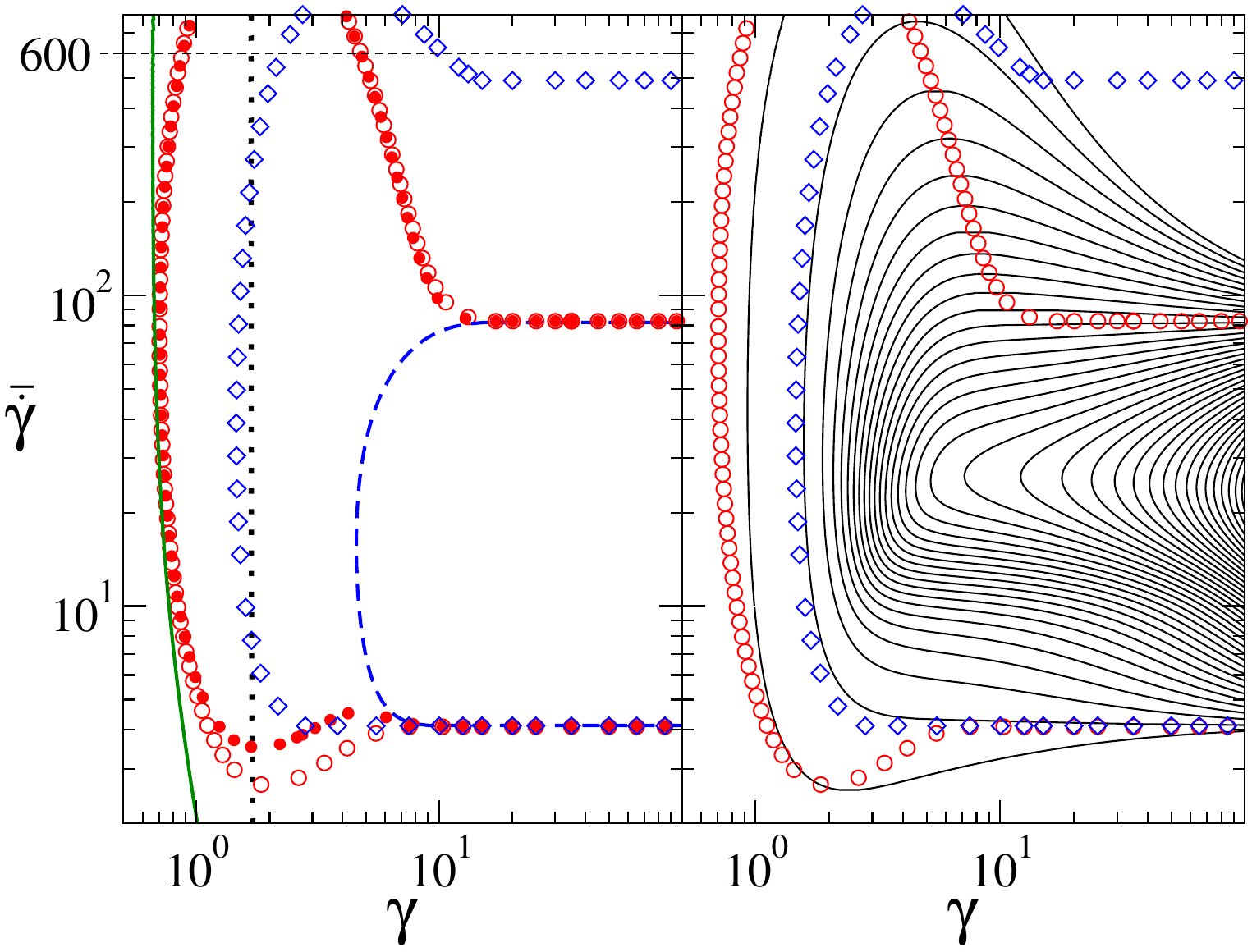}
 \caption{Shear startup in the nRP model with a non-monotonic
   constitutive curve: $\beta = 0.4$, $\eta = 10^{-4}$. {\bf Left
     panel.}  Black dotted line: location of stress overshoot. Solid
   line: location of onset of `elastic instability' (\ref{eqn:
     crit_elas_2dof}). Dashed line encloses region of of `viscous
   instability' (\ref{eqn: crit_viscous}). Large open circles enclose
   region of linear instability according to full criterion (\ref{eqn:
     crit_default}). Small closed circles enclose region of linear
   instability according to the criterion with cross terms omitted
   (\ref{eqn: no_cross}).  Diamonds enclose region in which
   significant shear banding is seen in our spatially aware nonlinear
   simulations.  {\bf Right panel.} Solid lines: contour lines of
   equal $|\delta\gdot_{n=1}|/\gdotbar = 10^{M}$ for integer
   $M$ found by directly integrating the linearised
   Eqns.~\ref{eqn: crit3}).  (First contour: $M = -2$ and we show only
   contours $M\geq-2$.)  Circles and diamonds as in left panel.}
 \label{fig: portrait_noiseinitonly_beta0.4}
\end{figure}
\fi

\iffigures
\begin{figure}[tbp]
 \includegraphics[width=8.5cm]{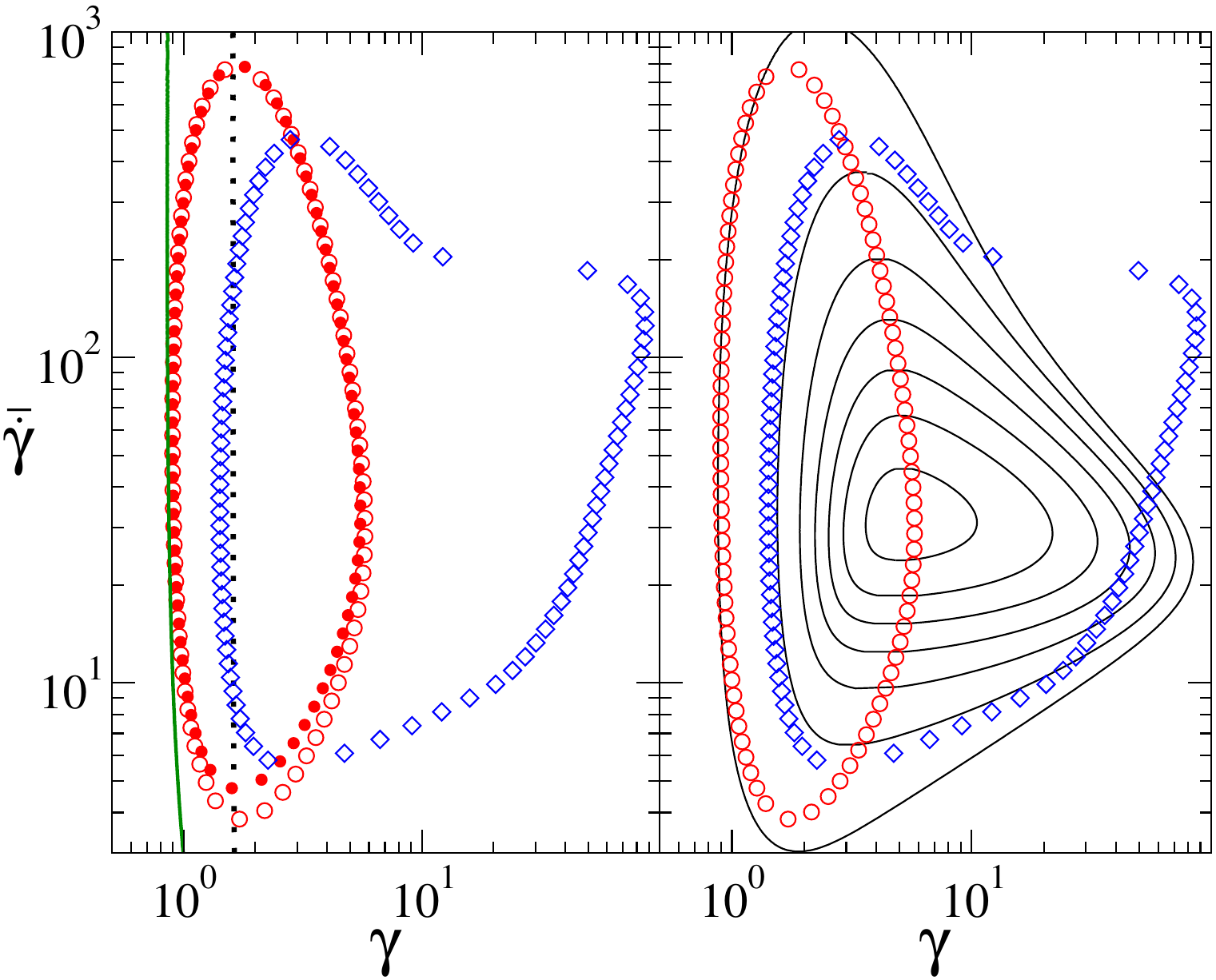}
 \caption{ As in Fig.~\ref{fig: portrait_noiseinitonly_beta0.4} but now
   for nRP model with a monotonic constitutive curve, $\beta = 1.0$,
   The `viscous instability' is absent in this case, but the `elastic
   instability' remains.  }
 \label{fig: portrait_noiseinitonly_beta1}
\end{figure}
\fi

As can be seen, the full criterion~(\ref{eqn: crit_default}) is very
well represented by the much simpler elastic one~(\ref{eqn:
  crit_elas_2dof}) across a wide range of shear rates during the early
stage of shear startup towards the left hand side of the
$\gdot_0,\gamma_0$ plane, and by the even simpler viscous
one~(\ref{eqn: crit_viscous}) at the far right hand side. This ability
of the `elastic' and `viscous' criteria separately to capture the full
criterion in these regimes leads us further to indicate by the small
full circles the region in which a criterion formed simply by summing
the elastic terms ~(\ref{eqn: crit_elas_2dof}) and the viscous terms
~(\ref{eqn: crit_viscous}) is met:
\beqn
\label{eqn: no_cross}
\partial_{\gdot_0}\base{\Sigma}|_{\base{\gamma}}-\frac{1}{\gdot_0^2}\text{tr}\tens{M}\,
\partial_{\gamma_0}\base{\Sigma}|_{\base{\gdot}} +
\frac{1}{\gdot_0}\partial_{\gamma_0}^{2}\,\base{\Sigma}|_{\base{\gdot}}
< 0.  \eeqn
This simple criterion performs well in capturing the region of
instability across the full plane of $(\gamma_0,\gdot_0)$, apart a
small region at the bottom left.

As just described, the left panel of Fig.~\ref{fig:
  portrait_noiseinitonly_beta0.4} concerns our criteria for the onset
of a positive eigenvalue of the linearised system of
equations~(\ref{eqn: crit3}), which we propose indicates the onset of
a linear instability to shear banding as the underlying homogeneous
base state evolves in time.  However, as noted above, the concept of a
time-dependent eigenvalue should be treated with some caution.
Therefore in the right hand panel of Fig.~\ref{fig:
  portrait_noiseinitonly_beta0.4} we show results obtained by
integrating the linearised equations~(\ref{eqn: crit3}) directly.  The
solid lines are contours of equal $|\delta\gdot_{n=1}|$, obtained by
this process of integration. As can be seen the region of growth and
decay in this heterogeneous perturbation agrees well with the
eigenvalue-based criteria in the left subpanel, confirming that our
concept of a time-dependent eigenvalue is indeed useful.

We also summarise in Fig.~\ref{fig: portrait_noiseinitonly_beta0.4}
the results of a series of fully nonlinear spatially-aware simulations
of shear startup, performed for a wide range of values of $\gdotbar$
at closely spaced intervals. Again, any horizontal slice across this
plane corresponds to one of these runs at a given $\gdot_0$. The
diamonds show the region of this plane of strain-rate and strain for
which significant shear banding is observed. (We choose
$\Delta_{\gdot} > 0.05\gdotbar$ as a criterion for significant
banding.)  As can be seen, the region of significant banding agrees
well with expectations based on the linear calculation alone in most
regions of the plane. However a window of shear rates either side of
the regime of viscous linear instability deserves further comment.
Here the nonlinear simulations remain significantly banded in steady
state, even though the linear system has returned to stability by
then.  For such shear rates, a state of homogeneous shear on the
underlying constitutive curve is indeed linearly stable, but in fact
only metastable: the true steady state is banded.  Finally at very
high shear rates {\it e.g.}  $\gdotbar = 600$, we observe shear bands
that form transiently in startup, triggered by the `elastic'
instability, but that return to homogeneous flow in steady state.

In summary, the overall stability portrait of the nRP model with a
nonmonotonic constitutive curve comprises, in this plane of
strain-rate and strain, a vertical patch of `elastic instability' at
the left side of the plane, and a horizontal patch of `viscous
instability' at the right hand side. In between these limits there is
a continuous cross-over between the two instabilities.

In contrast, for model parameters for which the constitutive curve is
monotonic, the patch of `viscous instability' is absent and the
eventual steady flowing state is homogeneous at all strain rates.
Importantly, however, a patch of `elastic instability' remains with
onset at a strain $\gamma\approx 1$, again closely associated with the
startup stress overshoot at strain $\gamma\approx 1.7$.  See
Fig.~\ref{fig: portrait_noiseinitonly_beta1}.  This triggers
pronounced shear banding during startup, which however persists only
transiently, decaying at larger strains to leave homogeneous flow in
steady state.  A single startup run corresponding to a horizontal
slice across this plane at $\gdot_0=30.0$ is explored in~\ref{fig:
  RP_example_transient} (top row). In steady state, the flow is
homogeneous with a stress value indicated by the upper cross in
Fig.~\ref{fig: RP_constitutive_curves}.

\iffigures
\begin{figure}[tbp]
 \includegraphics[width=6.0cm,height=6.0cm]{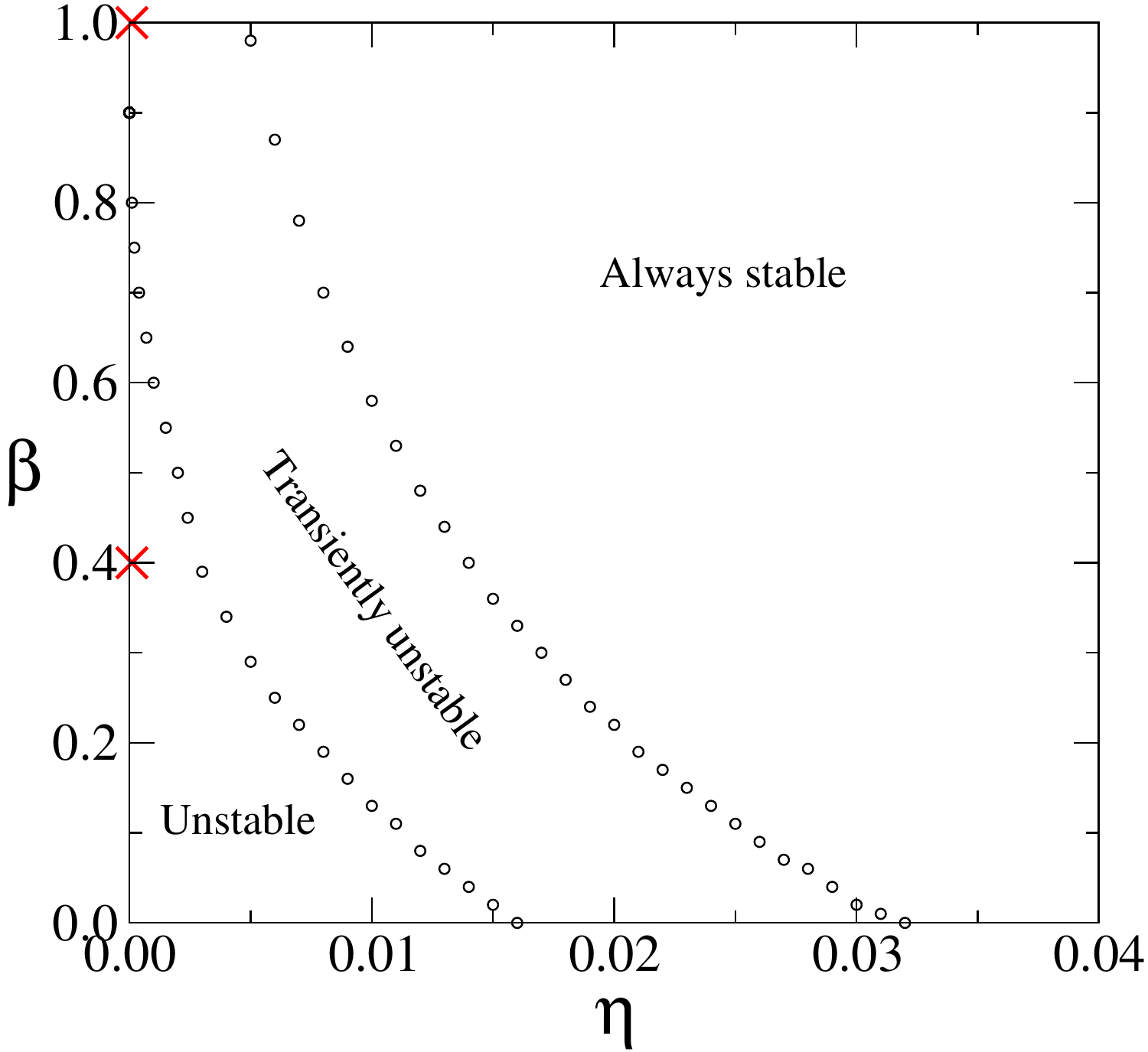}
 \caption{Summary of the linear stability properties of the nRP model
   as a function of the model parameters $\beta,\eta$, during shear
   startup at a shear rate in the region of minimum slope of the
   constitutive curve. \emph{`Unstable'}: linearly unstable to shear
   heterogeneity at steady state. \emph{`Transiently unstable'}:
   system shows linear instability at some time during shear startup
   but returns to linearly stable at steady state. \emph{`Always
     stable'}: the system is always linearly stable to shear
   heterogeneity. Crosses `$\times$' at: $\beta = 1, 0.4$ at $\eta =
   10^{-4}$ indicate the two sets of parameter values explored in
   detail in the text.}
 \label{fig: RP_MI_region}
\end{figure}
\fi

So far, we have explored in detail one set of model parameters
$(\beta,\eta)=(0.4, 10^{-4})$ for which the underlying constitutive
curve is non-monotonic and a viscous instability persists to steady
state; and one set of parameters $(\beta,\eta)=(1.0,10^{-4})$ for
which the constitutive curve is monotonic and shear bands form only
transiently. Denoting these two distinct cases by ``unstable'' and
``transiently unstable'' respectively, we summarise the model's
behaviour in the full plane of $(\beta,\eta)$ in Fig.~\ref{fig:
  RP_MI_region}.

\subsubsection{Stretching RP model}

\iffigures
\begin{figure}[tbp]
  \centering
	\includegraphics[width=7cm]
{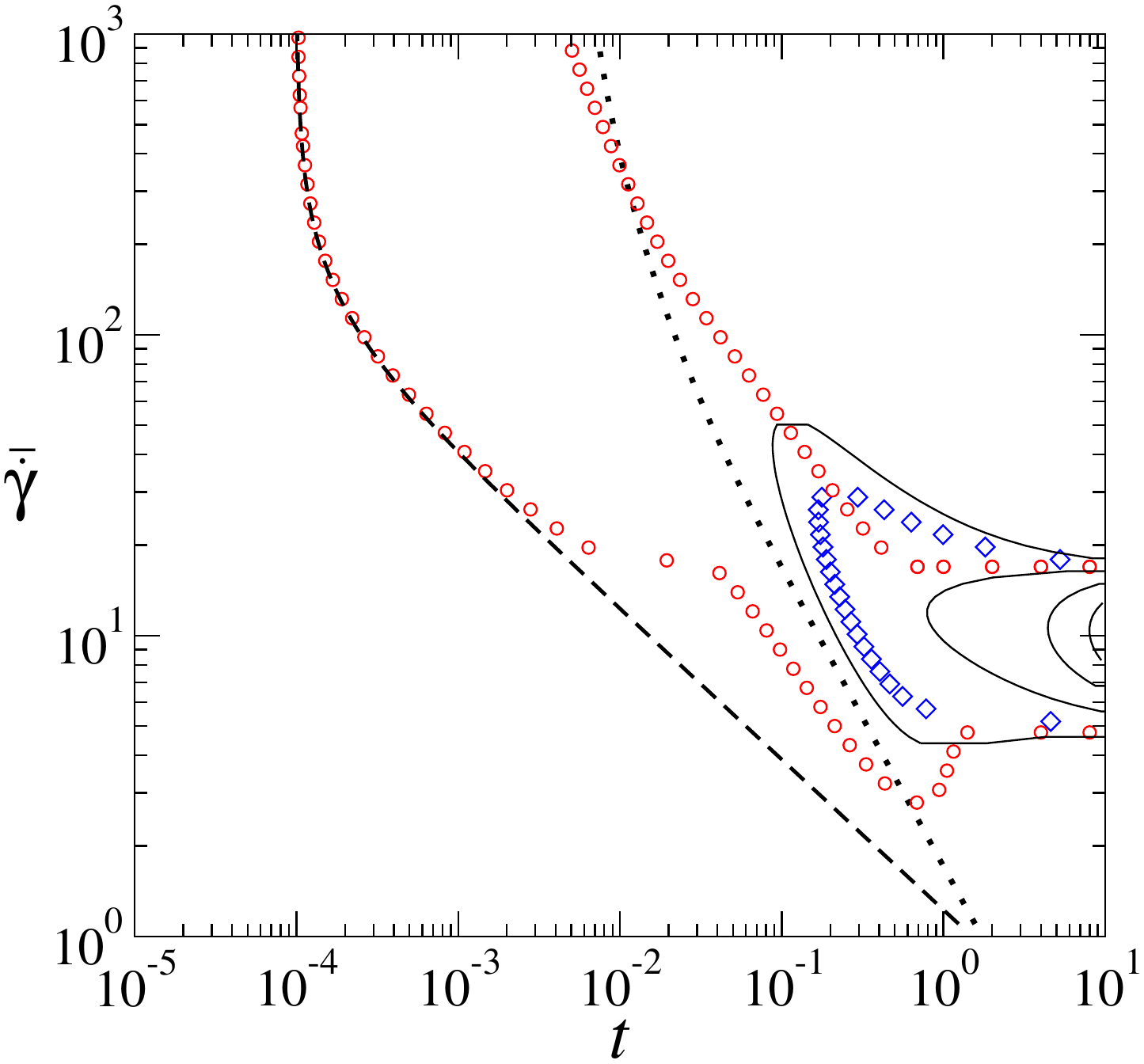}
\caption{Stability portrait of the sRP model in the plane of
  strain-rate and strain for model parameter values $\beta = 0.4$,
  $\eta = 10^{-4}$, $\taur = 10^{-2}$, $q = 5\times10^{-3}$. Solid
  lines show contours of equal $|\delta \gdot_{n}| = \gdot 10^M$ for
  integer $M$ found by directly integrating the linearised
  Eqns.~\ref{eqn: crit3}. (The contour nearest $\partial_{t\,}\Sigma =
  0$ has $M = -2$, and we show only contours $M \geq -2$.) Diamonds
  show the region of significant banding in a full nonlinear spatially
  aware simulation.  For shear rates in the non-stretching regime
  $\invtaud \ll \gdot \ll \invtaur$ we recover the behaviour discussed
  previously in the nRP model. In contrast an `sRP-specific'
  instability is seen in upper-left region of the plane with onset
  given by the formula in Eqn.~\ref{eqn: srpspecific}, which is shown
  by a dashed line.  However it does not precipitate significant
  banding. }
\label{fig: sRP_heterogeneous_1}
\end{figure}
\fi

In the previous subsection we discussed the predictions of the
non-stretch rolie-poly (nRP) model, in which any possibility of chain
stretch is switched off by setting $\taur=0$. Recall Eqns.~\ref{eqn:
  nRP_components}. We now turn to the sRP model in which chain stretch
is accounted for. Recall Eqns.~\ref{eqn: sRP_components}.

For applied strain rates $\gdot \ll \invtaur$, no appreciable stretch
arises even in the sRP model, and the nRP results discussed above
apply directly. This can be seen in Fig.~\ref{fig:
  sRP_heterogeneous_1}: the dynamics in the regime $\gdot\ll
\taur^{-1}$ is the same as discussed above for the nRP model.

The focus in this section is therefore on shear startup runs performed
at shear rates $\gdot > \invtaur$, for which appreciable chain stretch
does develop. Here the system exhibits an early-time ($t<\taur$)
stress-strain behaviour corresponding to that of a linear elastic
solid, with $\Sigma = G\gamma$. At longer times $t> \taur$ the
relaxation of chain stretch leads to deviation from this linear
relation and a stress signal that decreases with strain, after an
overshoot. In contrast to the nRP model, in which stress overshoot
occurs at a fixed value of the strain, in the stretching regime this
overshoot now occurs at a fixed time $t \sim \taur$. Accordingly, the
stress startup curve no longer converges to a limiting function of
strain $\Sigma=\Sigma(\gamma)$ at high shear rates, and the concept of
a purely `elastic instability' no longer applies. As can be seen in
Fig.~\ref{fig: sRP_heterogeneous_1} the onset of an elastic
instability just before overshoot is apparent only in the non-stretch
regime $\gdot \ll \invtaur$ explored previously in the nRP model, and
breaks down for $\gdot > \invtaur$.

Surprisingly, however, we find a new linear instability, specific to
the stretching regime $\gdot_0 > \invtaur$, that sets in at a time
{\em before} overshoot given by
\beqn
t_s = \frac{3}{2\taud\base{\gdot}^2} + \eta/G \quad \text{ for } \base{\gdot} \gg \invtaur.
\label{eqn: srpspecific}
\eeqn
However this instability disappears again even before the overshoot
occurs and never leads to observable banding, so we pursue it no
further here.

\subsubsection{Rolie-poly model: relation to shear startup  experiments}

We have shown that the RP model shows rich time-dependent banding
dynamics during shear startup in the nonstretching regime $\invtaud
\ll \gdot \ll \invtaur$.  These results, together with those of Adams
\etal \cite{Adamsetal2011, Adamsetal2009a, Adamsetal2009b},
demonstrate that the RP model captures the experimental phenomenology
of entangled polymeric fluids in this shear startup protocol. We
summarise this now, divided into three classes (i) - (iii) for
convenience.

(i) For imposed shear rates in the negatively sloping regime of a
non-monotonic underlying constitutive curve, we find shear banding
that sets in around the time of an overshoot in the stress startup
curve and is initially sufficiently violent as to lead to elastic
recoil and negative local shear rates or velocities. It persists to
steady state but with much smaller magnitude than around the time of
overshoot.  This is consistent with experimental observations in
Refs.~\cite{Wangetal2008a, Wangetal2008b, Wangetal2009a,
  Wangetal2006b, Wangetal2009d}. (ii) For imposed shear rates some
distance above the negatively sloping regime of a non-monotonic
constitutive curve, we again find violent shear banding setting in
around the time of stress overshoot during startup. However these
bands persist only transiently, and decay to leave a homogeneous
steady state. This is consistent with experimental observations in
Refs.~\cite{Wangetal2009a,Wangetal2008a}.  (iii) For a monotonic
constitutive curve we again see pronounced transient banding triggered
by stress overshoot, which decays to leave a homogeneous steady state,
as in experimental observations in
Refs.~\cite{Huetal2007a,Wangetal2008a,Wangetal2009a}.

The main new contribution of the present manuscript has been to place
these observations in the context of our general analytical criterion
for the onset of banding.

\subsection{Numerical results: Giesekus model}
\label{section: giesekus_startup}

\iffigures
\begin{figure}[tbp]
  \includegraphics[width=8cm]{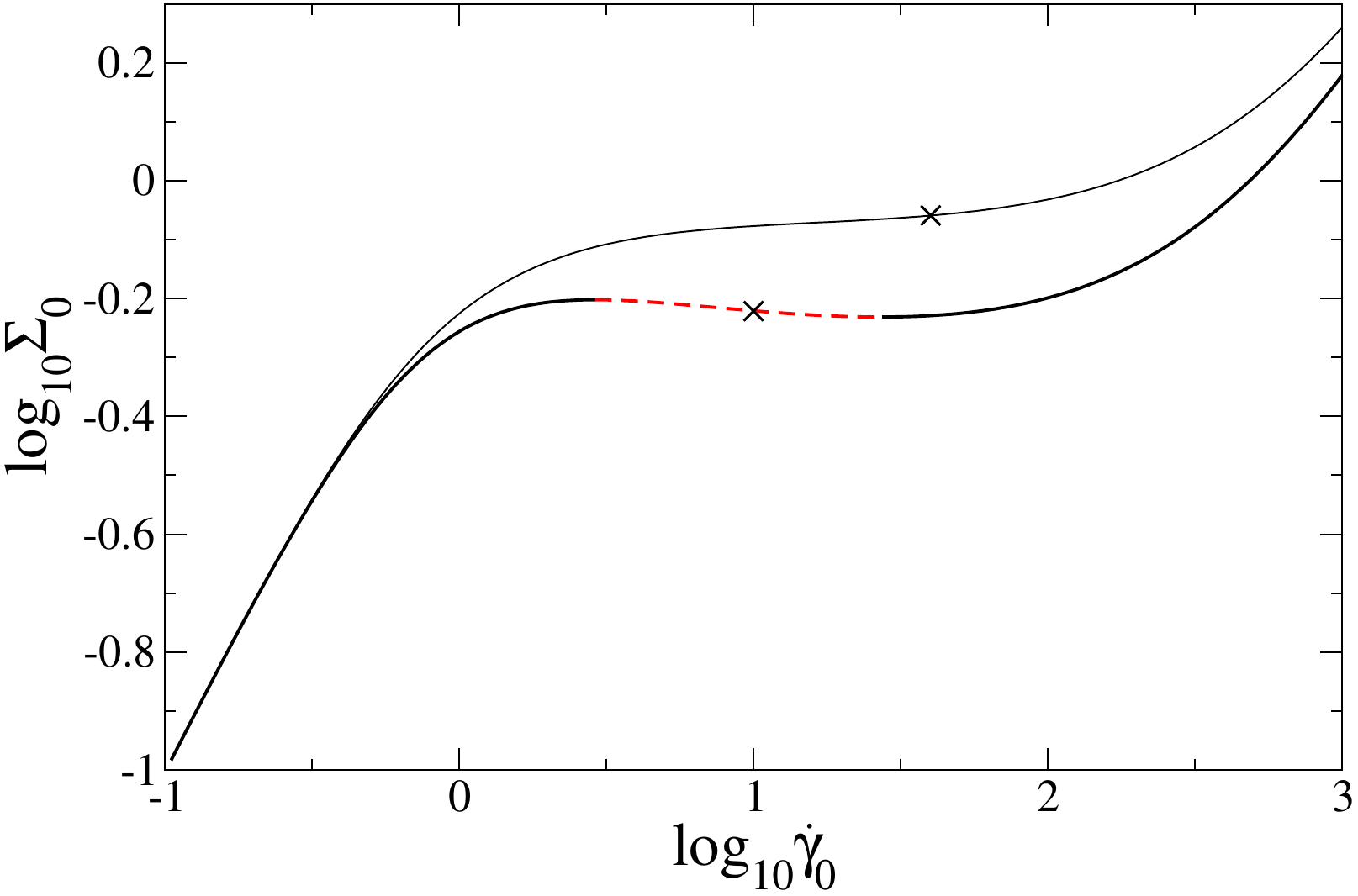}
  \caption{Constitutive curves for the Giesekus model with $\alpha =
    0.6, 0.8$ (top to bottom) and $\eta = 10^{-3}$. Regime of linear
    instability is shown as a dashed line. Crosses indicate shear
    rates $\gdot = 10, 40$ for which time-dependent behaviour is
    shown in Figs.~\ref{fig: Giesekus_example_transient1} and
    \ref{fig: Giesekus_example_transient3} respectively.}
  \label{fig: constitutive_curve_giesekus}
\end{figure}
\fi

We now discuss our numerical results for shear startup in the Giesekus
model. Our aim is to address whether this model is capable of
capturing the time-dependent shear banding behaviour observed
experimentally in entangled polymers, summarised in (i) - (iii) above.
To provide a fair comparison with the RP model, we choose values of
the parameter $\alpha$ giving constitutive curves that are as closely
comparable between the models as possible. Compare Fig.~\ref{fig:
  constitutive_curve_giesekus} with Fig.~\ref{fig:
  RP_example_transient}.

To explore class (i) behaviour, we consider the nonmonotonic
constitutive curve of Fig.~\ref{fig: constitutive_curve_giesekus} and
perform a shear startup at a value of the shear rate represented by
the cross in the negatively sloping regime. See Fig.~\ref{fig:
  Giesekus_example_transient1}. The stress startup curve closely
resembles that of the RP model, with a pronounced overshoot.  However
the Giesekus model apparently lacks the region of pronounced elastic
instability associated with this overshoot. Instead, the degree of
banding $\Delta_{\gdot}$ rises monotonically and only becomes
significant at long times, when the criterion
$\partial_{\gdot}\Sigma|_{\gamma \to \infty}<0$ for viscous
instability and steady state banding is met. For shear rates outside
the negatively sloping regime of this constitutive curve (not shown),
we find no banding during startup or in steady state. The Giesekus
model therefore fails to address classes (i) and (ii) of polymeric
startup behaviour described above.

\iffigures
\begin{figure}[tbp]
	\includegraphics[width=8.5cm,height=8.5cm]{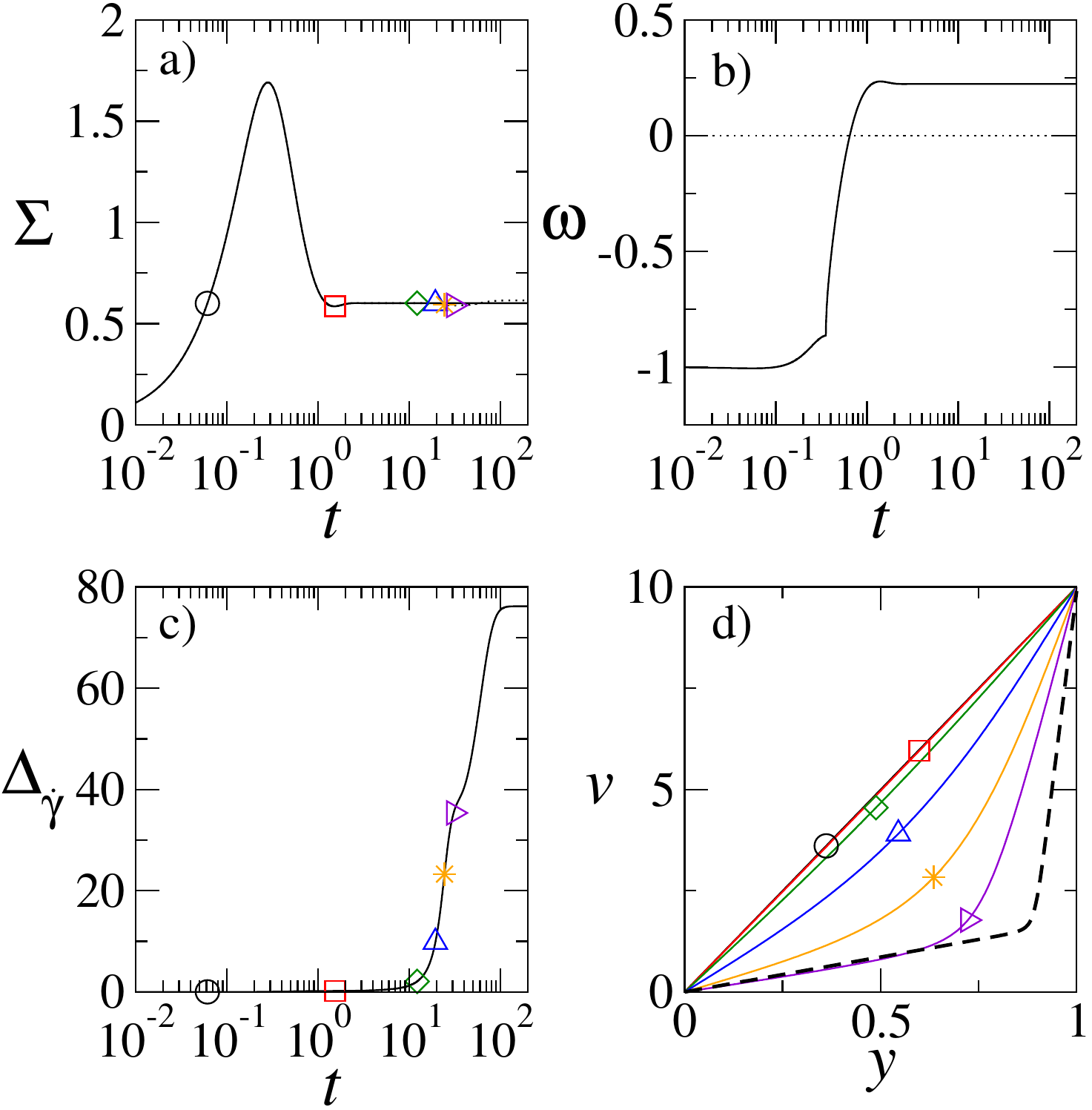}
	\caption{Shear startup in Giesekus model at an applied shear
          rate $\gdotbar=10$ in the negative sloping regime of the
          non-monotonic constitutive curve of Fig.~\ref{fig:
            constitutive_curve_giesekus}. (a) Shear stress startup
          curve (results with heterogeneity allowed are
          indistinguishable from the homogeneously constrained
          system). (b) Largest real part of any eigenvalue from linear
          stability analysis $\omega$. (c) Degree of banding in the
          nonlinear simulation, $\Delta_{\gdot} = \gdot_{\rm
            max}-\gdot_{\rm min}$.  (d) Snapshots of the velocity
          profile in the nonlinear simulation at strains corresponding
          to symbols in (a, c).  The steady state velocity profile is
          shown as a thick, dashed line.  Magnitude of initial noise:
          $q=10^{-2}$.  }
	\label{fig: Giesekus_example_transient1}
\end{figure}
\fi

\iffigures
\begin{figure}[tbp]
	\includegraphics[width=8.5cm,height=8.5cm]{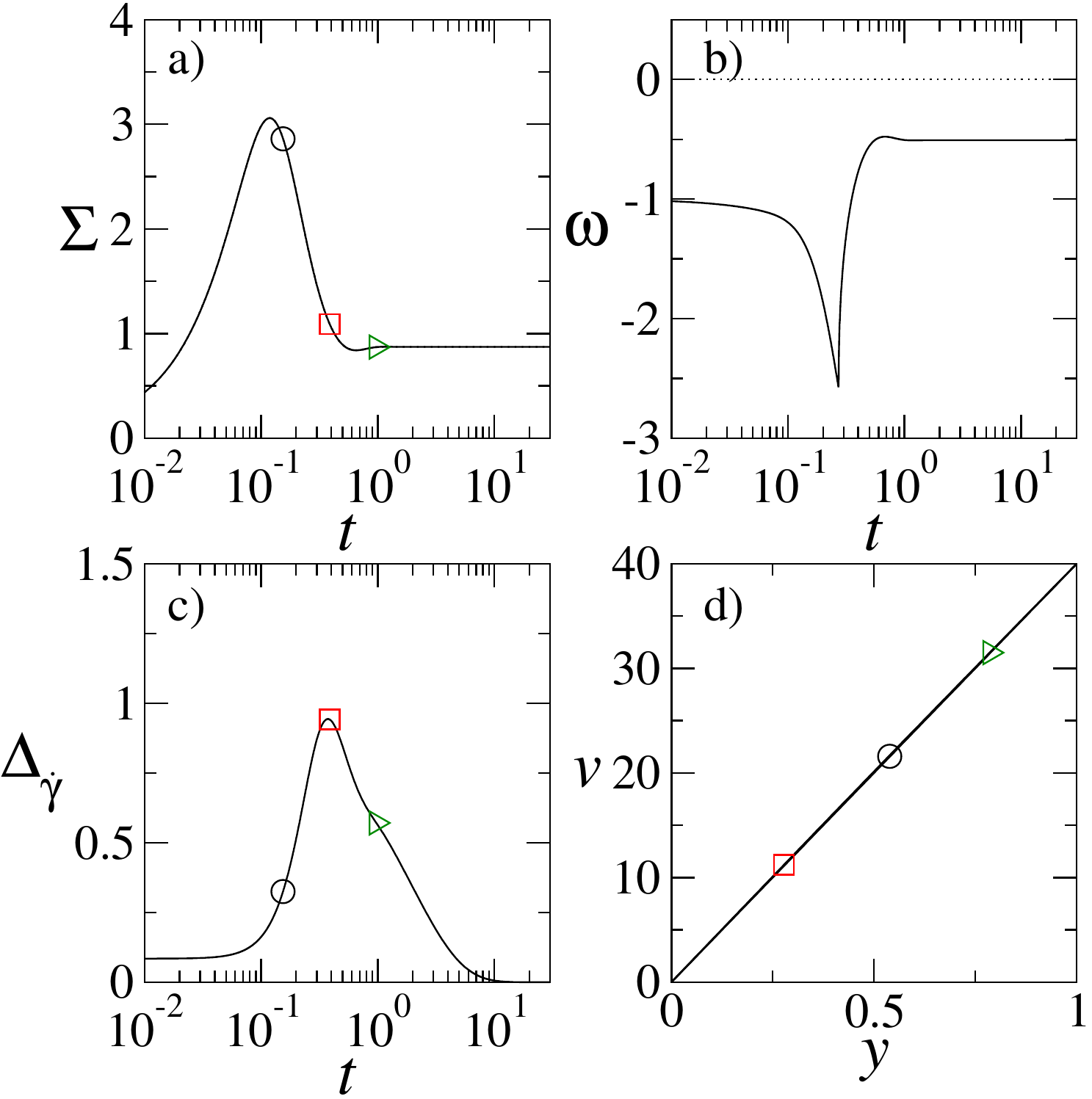}
	\caption {As in Fig.~\ref{fig: Giesekus_example_transient1},
          but now for an applied shear rate $\gdotbar = 10$ at the
          weakest slope of the constitutive curve of Fig.~\ref{fig:
            constitutive_curve_giesekus} with $\alpha = 0.8,\eta =
          10^{-3}$.}
	\label{fig: Giesekus_example_transient3}
\end{figure}
\fi

To explore class (iii) behaviour, we consider the monotonic
constitutive curve of Fig.~\ref{fig: constitutive_curve_giesekus} and
perform a shear startup at a value of the shear rate represented by
the cross in the region of weakest slope.  See Fig.~\ref{fig:
  Giesekus_example_transient3}. The stress startup curve again closely
resembles that of the RP model. However the magnitude of transient
shear banding is significantly diminished in comparison, never
exceeding $10\%$ of the overall imposed shear rate $\gdotbar$.
Indeed, no eigenvalue shows a positive real part during startup for a
monotonic constitutive curve in this model.  Accordingly, the Giesekus
model lacks the pronounced `elastic instability' of the RP model and
fails to address class (iii) behaviour also.

A thorough exploration of this parameter space $\alpha,\eta,\gdotbar$
(not shown) confirms that the above comments of negligible banding
during startup apply generically in the Giesekus model. This obviously
constrasts sharply with our results for the RP model above. The reason
for this appears to be the different structure of the loading terms
between the Giesekus and RP models.  Compare Eqns.~\ref{eqn:
  Giesekus_components} with Eqns.~\ref{eqn: nRP_components}. Because
of this difference, the Giesekus model does not attain a limiting
nonlinear shear startup curve $\Sigma(\gamma)$ at high shear rates,
and accordingly lacks the possibility of an elastic instability.

\section{Conclusions and outlook}
\label{section: conclusions}

We have explored theoretically the onset of shear banding in polymeric
and wormlike micellar fluids for three of the most common
time-dependent rheological protocols: step stress, strain ramp, and
shear startup.  For each protocol we have developed a fluid-universal
criterion for the onset of linear instability to shear banding. We
have supported these predictions with numerical simulations of the
rolie-poly and Giesekus models. Between these models, we have found
the rolie-poly model to be effective in capturing the observed
experimental phenomenologym. In contrast, the Giesekus model apparently
fails to do so.

Following the imposition of a step stress, a base state of initially
homogeneous creep response becomes unstable to the formation of shear
bands during any regime in which the shear rate $\gdot(t)$ is
simultaneously upwardly curving and upwardly sloping in time
$\partial_{t\,}^2\gdot /
\partial_{t\,}\gdot>0$. We believe this criterion to be universal in
all models for the rheology of complex fluids of which we are aware.
We showed that such a regime does indeed arise in both the Giesekus
and RP models for imposed stresses nearest those on the weakest slope
of the underlying constitutive curve of shear stress as a function of
shear rate. However the magnitude of the resulting shear banding only
attains a magnitude consistent with experimental findings
\cite{Wangetal2008b, Wangetal2003a, Huetal2008a, Wangetal2008c,
  Huetal2005a, Wangetal2009d} in the RP model.

For the strain ramp protocol, a base state of initially homogeneous
shear response is left unstable immediately post-ramp if the stress
had been decreasing with strain towards the end of the ramp.  We
believe this criterion to be general for all ramps applied at a rate
exceeding the inverse of the material's intrinsic relaxation time, for
any viscoelastic constitutive equation that can be expressed as the
sum of separate loading and relaxation terms.

In the RP model, we demonstrated numerically that this criterion for
instability immediately post-ramp is met for ramp rates in the
nonstretching regime $\invtaud \ll \gdot_0 \ll \invtaur$, and ramp
amplitudes $\gamma_0 \gtrsim 1.7$. However we further explored the RP
model's full relaxation as a function of the time elapsed since the
ramp ended, following ramps that are either slow or fast relative to
the rate of chain stretch relaxation $\invtaur$.  In the absence of
convective constraint release (CCR), the stress relaxation function of
a `fast' ramp drops onto that of a `slow' ramp once chain stretch has
relaxed.  This leads to a delayed shear banding instability following
a `fast' ramp, even though in that case the stress increased
monotonically with strain during the ramp.  In contrast CCR tends to
stabilise the system against this `delayed' banding instability.  In
capturing such rich phenomenology, we again find the RP model capable
of addressing the experimental data for polymeric fluids, while the
Giesekus model would be expected to perform poorly in comparison.

Finally we explored the onset of shear banding in the shear startup
protocol. For materials that attain a limiting nonlinear startup curve
of stress as a function of strain at high strain rates, we identified
separate `elastic' and `viscous' instabilities that respectively act
at early and late times during startup. We confirmed the presence of
these two distinct regimes in a numerical study of the RP model, which
shows a violent elastic instability at early times during startup at
rates $\invtaud \ll \gdot \ll \invtaur$, closely associated with an
overshoot in the stress startup signal. This banding persists to
steady state in any regime of negative slope in the underlying
constitutive curve ({\it i.e.} of viscous instability), but with a
`degree of banding' that is much weaker than that seen during the
initial elastic instability.  In contrast the Giesekus model does not
attain a limiting startup curve of stress as a function of strain at
high strain rates and lacks a violent elastic instability during
startup, in contrast to experimental observations. It does, however,
correctly capture steady state banding.  Accordingly we conclude that
the RP model provides a good description of shear banding during
time-dependent flows in entangled polymeric fluids, while the Giesekus
model performs poorly in comparison.

The reader is referred to a separate manuscript in which the
fluid-universal criteria that we have derived here (and discussed in
detail in the context of polymer fluids) are explored in the context of a
broad class of disordered soft glassy materials including foams, dense
emulsions and colloids~\cite{SuzanneInProgress}.

\section{Acknowledgements}

The authors thank Stephen Agimelen, Peter Olmsted, and Richard Graham for helpful discussions, and the UK's EPSRC (EP/E5336X/1) for funding.



\end{document}